\pdfoutput=1
\documentclass[twocolumn,svgnames]{aastex63}
\usepackage{CJKutf8}

\usepackage{amsmath,amsthm,amssymb}
\usepackage{graphicx}
\usepackage{svg}
\usepackage{microtype}
\usepackage{needspace}
\usepackage{hyperref}
\usepackage{xcolor}
\nonstopmode

\newcommand{\uat}[2]{\href{http://astrothesaurus.org/uat/#2}{#1 (#2)}}
\newcommand{\gaia}{\textit{Gaia}}
\newcommand{\w}[2]{\ensuremath{{#1}~{\rm {#2}}}}
\newcommand{\po}{\Phi_0}
\DeclareMathOperator{\atantwo}{atan2}

\begin{document}
\begin{CJK*}{UTF8}{gbsn}

  %%%%%%%%%%%%%%%%%%%%%%%%%%%%%%%%%%%
  \title{The Mass of the Milky Way from the H3 Survey}

  \correspondingauthor{Jeff Shen}
  \email{jshenschool@gmail.com}

  \author[0000-0001-6662-7306]{Jeff Shen}
  \affiliation{Canadian Institute for Theoretical Astrophysics, University of Toronto, 60 St. George Street, Toronto, ON M5S 3H8, Canada}
  \affiliation{David A. Dunlap Department of Astronomy \& Astrophysics, University of Toronto, 50 St. George Street, Toronto, ON M5S 3H4, Canada}
  \affiliation{Department of Statistical Sciences, University of Toronto, 100 St George Street, Toronto ON, M5S 3G3, Canada}

  \author[0000-0003-3734-8177]{Gwendolyn M. Eadie}
  \affiliation{David A. Dunlap Department of Astronomy \& Astrophysics, University of Toronto, 50 St. George Street, Toronto, ON M5S 3H4, Canada}
  \affiliation{Department of Statistical Sciences, University of Toronto, 100 St George Street, Toronto ON, M5S 3G3, Canada}

  \author{Norman Murray}
  \affiliation{Canadian Institute for Theoretical Astrophysics, University of Toronto, 60 St. George Street, Toronto, ON M5S 3H8, Canada}
  \affiliation{Canada Research Chair in Theoretical Astrophysics}

  \author[0000-0002-5177-727X]{Dennis Zaritsky}
  \affiliation{Steward Observatory, University of Arizona, 933 North Cherry Avenue, Tucson, AZ 85721-0065, USA}

  \author[0000-0003-2573-9832]{Joshua S. Speagle (沈佳士)}
  \affiliation{David A. Dunlap Department of Astronomy \& Astrophysics, University of Toronto, 50 St. George Street, Toronto, ON M5S 3H4, Canada}
  \affiliation{Department of Statistical Sciences, University of Toronto, 100 St George Street, Toronto ON, M5S 3G3, Canada}
  \affiliation{Dunlap Institute for Astronomy and Astrophysics, University of Toronto, 50 St. George Street, Toronto ON, M5S 3H4, Canada}

  \author[0000-0001-5082-9536]{Yuan-Sen Ting (丁源森)}
  \altaffiliation{Hubble Fellow}
  \affiliation{Institute for Advanced Study, Princeton, NJ 08540, USA}
  \affiliation{Department of Astrophysical Sciences, Princeton University, Princeton, NJ 08540, USA}
  \affiliation{Observatories of the Carnegie Institution of Washington, 813 Santa Barbara Street, Pasadena, CA 91101, USA}
  \affiliation{Research School of Astronomy \& Astrophysics, Australian National University, Cotter Rd., Weston, ACT 2611, Australia}
  \affiliation{Research School of Computer Science, Australian National University, Acton ACT 2601, Australia}

  \author[0000-0002-1590-8551]{Charlie Conroy}
  \affiliation{Center for Astrophysics $|$ Harvard \& Smithsonian, 60 Garden Street, Cambridge, MA 02138, USA}

  \author[0000-0002-1617-8917]{Phillip A. Cargile}
  \affiliation{Center for Astrophysics $|$ Harvard \& Smithsonian, 60 Garden Street, Cambridge, MA 02138, USA}

  \author[0000-0002-9280-7594]{Benjamin D. Johnson}
  \affiliation{Center for Astrophysics $|$ Harvard \& Smithsonian, 60 Garden Street, Cambridge, MA 02138, USA}

  \author[0000-0003-3997-5705]{Rohan P. Naidu}
  \affiliation{Center for Astrophysics $|$ Harvard \& Smithsonian, 60 Garden Street, Cambridge, MA 02138, USA}

  \author[0000-0002-6800-5778]{Jiwon Jesse Han}
  \affiliation{Center for Astrophysics $|$ Harvard \& Smithsonian, 60 Garden Street, Cambridge, MA 02138, USA}

  \begin{abstract}
    The mass of the Milky Way is a critical quantity which, despite decades of research, remains uncertain within a factor of two. Until recently, most studies have \deleted{relied on}\added{used} dynamical tracers in the inner regions of the halo, relying on extrapolations to estimate the mass of the Milky Way. In this paper, we extend the hierarchical Bayesian model applied in \cite{Eadie2019} to study the mass distribution of the Milky Way halo; the new model allows for the use of all available 6D phase-space measurements. We use kinematic data of halo stars out to $142~{\rm kpc}$, obtained from the H3 Survey and \textit{Gaia} EDR3, to infer the mass of the Galaxy. Inference is carried out with the No-U-Turn sampler, a fast and scalable extension of Hamiltonian Monte Carlo. We report a median mass enclosed within $100~{\rm kpc}$ of $\rm M(<100 \; kpc) = 0.69_{-0.04}^{+0.05} \times 10^{12} \; M_\odot$ (68\% Bayesian credible interval), or a virial mass of $\rm M_{200} = M(<216.2_{-7.5}^{+7.5} \; kpc) =  1.08_{-0.11}^{+0.12} \times 10^{12} \; M_\odot$, in good agreement with other recent estimates. We analyze our results using posterior predictive checks and find limitations in the model's ability to describe the data. In particular, we find sensitivity with respect to substructure in the halo, which limits the precision of our mass estimates to $\sim 15\%$.
  \end{abstract}


  %%%%%%%%%%%%%%%%%%%%%%%%%%%%%%%%%%%

  \needspace{5\baselineskip}
  \section{Introduction}\label{sec:intro} %%%%%%%%%%%%%%%%%%%%%%%%%%%%%%%%%%%

  Understanding the dark matter halo of the Milky Way is critical for our understanding of the Galaxy. In particular, the mass of the dark matter halo---which dominates the mass of the Galaxy---as well as its profile, inform us about the dynamics and evolution of the Galaxy and that of its satellites \citep[e.g.,][]{Boylan-Kolchin2013, Kallivayalil2013}. The halo mass of a galaxy is also closely linked to its stellar mass \citep{More2009, Watson2013}, star formation, and quenching \citep{Behroozi2019}. Estimates of the dark halo mass are also critical for placing the Galaxy in a cosmological context.

Various efforts using a wide range of different methods have been made to constrain the mass of the dark halo. Commonly used methods include:
\begin{itemize}
  \item the timing argument \citep[e.g.,][]{Zaritsky1989, Li2008, Zaritsky2019},
  \item rotation curve analysis \citep[e.g.,][]{Rubin1970, Xue2008a, Nesti2013, Cautun2020, Karukes2020},
  \item escape velocities \citep[e.g.,][]{Smith2007, Piffl2014}, and 
  \item phase-space distribution functions \citep[e.g.,][]{Little1987, Eadie2015, Eadie2016, Eadie2017, Eadie2019, Li2020, Deason2021}. 
\end{itemize}
It is the last of these methods that we adopt in this paper.

The phase-space distribution function (DF) fully specifies a dynamical system through the 6D kinematic data of a tracer: three positions (observationally, these are R.A., Dec., and parallax) and three velocities (two proper motion components and a line-of-sight velocity). This makes it well-positioned to take advantage of recent proper motion data from \gaia\ EDR3 \citep{GaiaCollaboration2020, Lindegren2020}. A DF-based approach lends itself well to a probabilistic treatment, allowing for the use of flexible hierarchical Bayesian models. 

The use of Bayesian modeling in astronomy is not novel (see, e.g., \citealt{Little1987}, \citealt{Trotta2008}, and \citealt{Hilbe2017}); it been applied to many domains of astronomy \citep[e.g.,][]{Sale2012, Nicholl2017, Sestovic2018}. Most inference has relied on traditional random-walk Metropolis-Hastings algorithms \citep{Metropolis1953, Hastings1970}, Gibbs sampling \citep{Casella1992a}, or affine-invariant Markov chain Monte Carlo (MCMC) \citep{Goodman2010a, Foreman-Mackey2013}. 
However, new statistical developments have led to samplers that provide increased statistical and/or computational efficiency, especially for dealing with complex models and large amounts of data (i.e., high-dimensional problems) where classical MCMC algorithms \added{and ensemble methods }are prohibitively slow. 

Hamiltonian Monte Carlo (HMC) \citep{Duane1987a}, and in particular its extension called the No-U-Turn sampler (NUTS) \citep{Hoffman2014}, has been underutilized in astronomical research despite its promise of faster inference, possibly due to its need for differentiable models\added{ (automatic differentiation largely solves this problem)}. With HMC, inference for large models with thousands of parameters becomes tractable, opening up many possibilities. NUTS is implemented in the open-source Stan programming language \citep{Carpenter2017, StanDevTeam2018}, which is already widely used in ecology \citep[e.g.,][]{Authier2014, Stoddard2020}, epidemiology \citep[e.g.,][]{Lewis2017, Donnat2020}, cognitive science \citep[e.g.,][]{Jager2020, Leuker2020}, and time-series analysis \citep[e.g.,][]{Taylor2017}.

The primary contribution of this paper is the extension of the hierarchical Bayesian model for estimating the mass of the Galaxy, previously developed, tested, and applied by Eadie et al. \citep{Eadie2016, Eadie2017, Eadie2018, Eadie2019}.  
For this analysis, our new model has 1012 free parameters (4 parameters for the distribution function and 6 latent phase-space parameters for each of the 168 tracer objects)---this is intractable with the Gibbs sampler used in Eadie et al.'s code, called Galactic Mass Estimator (GME\footnote{\protect \url{https://github.com/gweneadie/GME}}).
Our second contribution is thus the application of an underutilized algorithm for Bayesian inference---HMC---which makes fitting such a model possible.

The layout of this paper is as follows. In Section \ref{sec:data} we describe the H3 dataset and the selection criteria for our sample.
In Section \ref{sec:model} we give an overview of our multilevel model.
In Section \ref{sec:inference} we explain Hamiltonian Monte Carlo (HMC) and the No-U-Turn sampler (NUTS), which we use to estimate our model parameters. 
In Section \ref{sec:discussion-mocks} we \deleted{demonstrate the speed of the new Stan code by applying it to two existing data sets that have already been analyzed in previous studies, and then }apply our new model to simulated data to determine the accuracy our mass estimates.
In Section \ref{sec:analysis} we obtain mass estimates using the real halo star data.
In Section \ref{sec:discussion} we investigate the robustness of our mass estimates by performing posterior predictive checks and by testing for systematic biases caused by substructure in the halo.
A summary of our findings is given in Section \ref{sec:conclusion}.

  % \needspace{5\baselineskip}
  \section{Observational Data}\label{sec:data} %%%%%%%%%%%%%%%%%%%%%%%%%%%%%%%%%%%

  Our data consist of 6D phase-space information of halo stars from the H3 survey \citep{Conroy2019}; these data will be made available in H3 DR1\footnote{We are using version V4.2.3.d20201031\_MSG of the catalog.}. The survey selects distant stars based on \gaia\ parallaxes and currently provides stellar parameters and spectrophotometric distances for roughly $140,000$ stars down to $r=18$, with more to come. 
Stellar parameters for the H3 survey were measured using the \verb|MINESweeper| code (for details, see \citealt{Cargile2020}), which estimates posteriors for all relevant parameters using the nested sampling code dynesty \citep{Speagle2020}.

We select our sample based on the criteria used in \cite{Zaritsky2019}, which are intended to remove stars with problematic parameter estimates:
\begin{itemize}
  \item spectral signal-to-noise ratio (SNR) of at least 3,
  \item stellar rotational velocity less than $5~{\rm km\,s^{-1}}$,
  \item effective temperature less than \w{7000}{K},
  \item absolute value of the radial velocity in the Galactic Standard of Rest less than $400~{\rm km\,s^{-1}}$,
  \item and distances from the Galactic center greater than 50 kpc.
\end{itemize}

With these criteria, we end up with a sample size of 168 stars out to \w{142}{kpc}. The parameters for these stars were estimated without a galactic density prior, and a flat distance prior from $1-200~{\rm kpc}$ was applied. 

\begin{figure*}[htbp!]
  \centering
  \gridline{%
    \fig{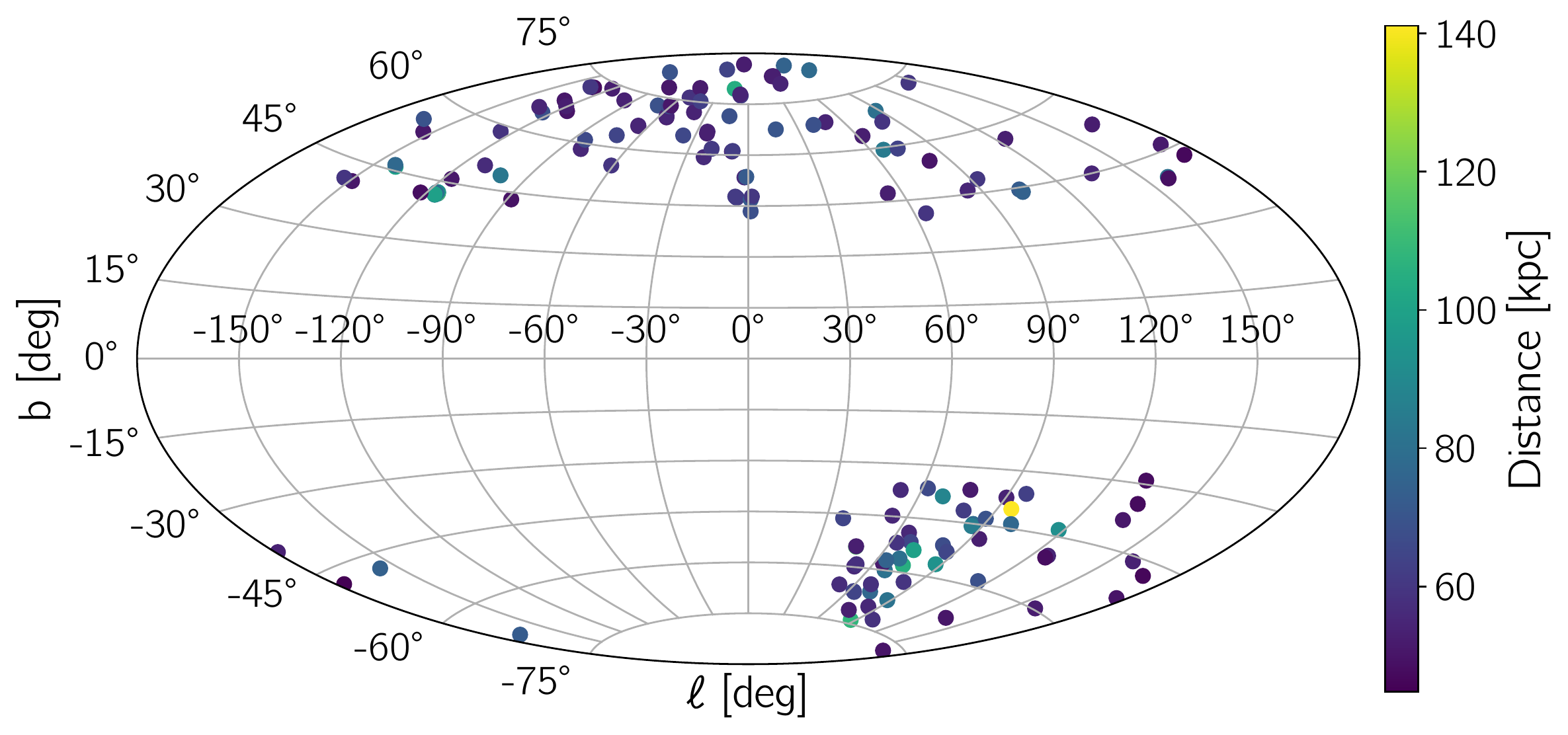}{0.57\textwidth}{(a)}
    \fig{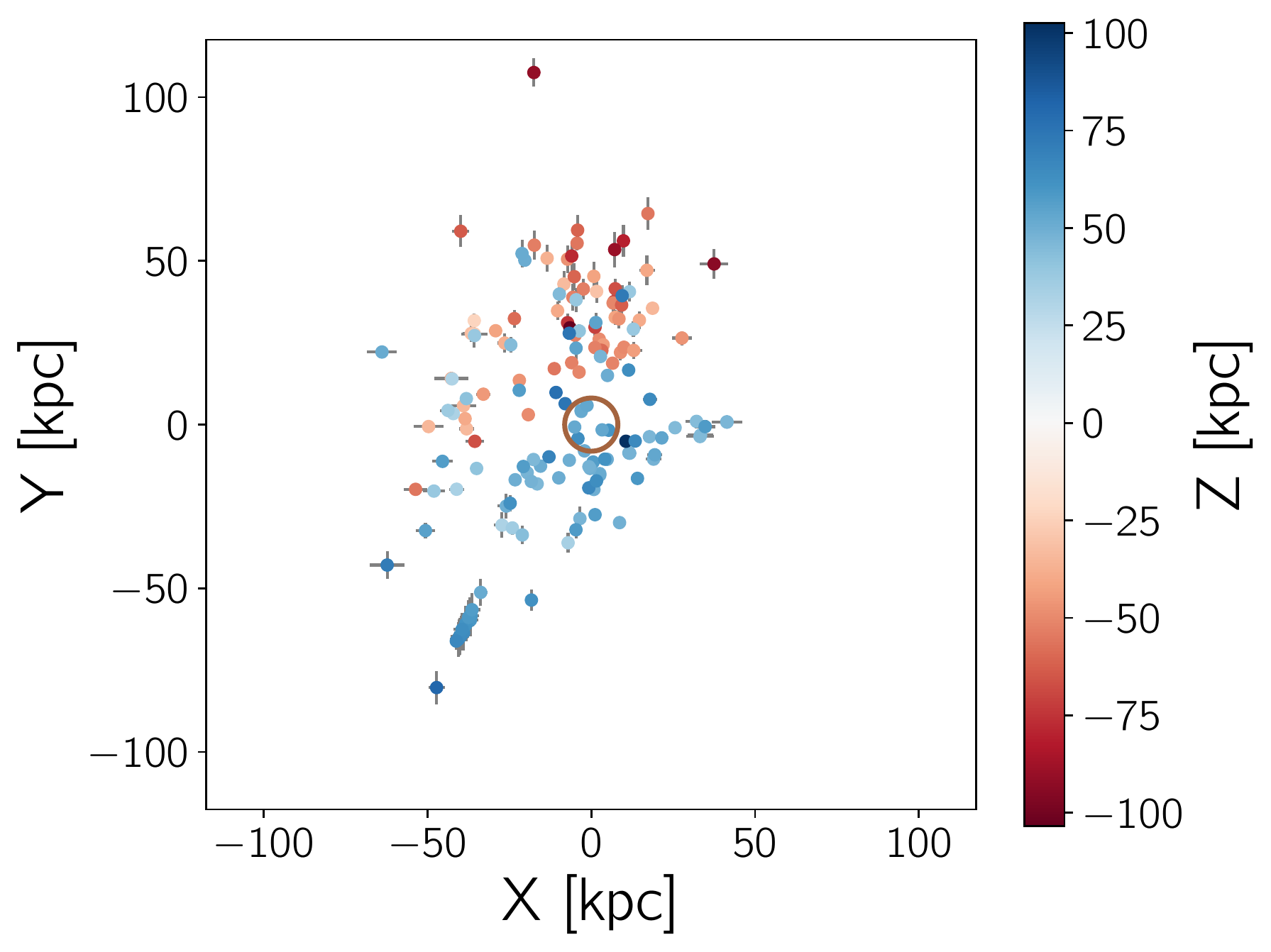}{0.37\textwidth}{(b)}
  }
  \gridline{%
    \fig{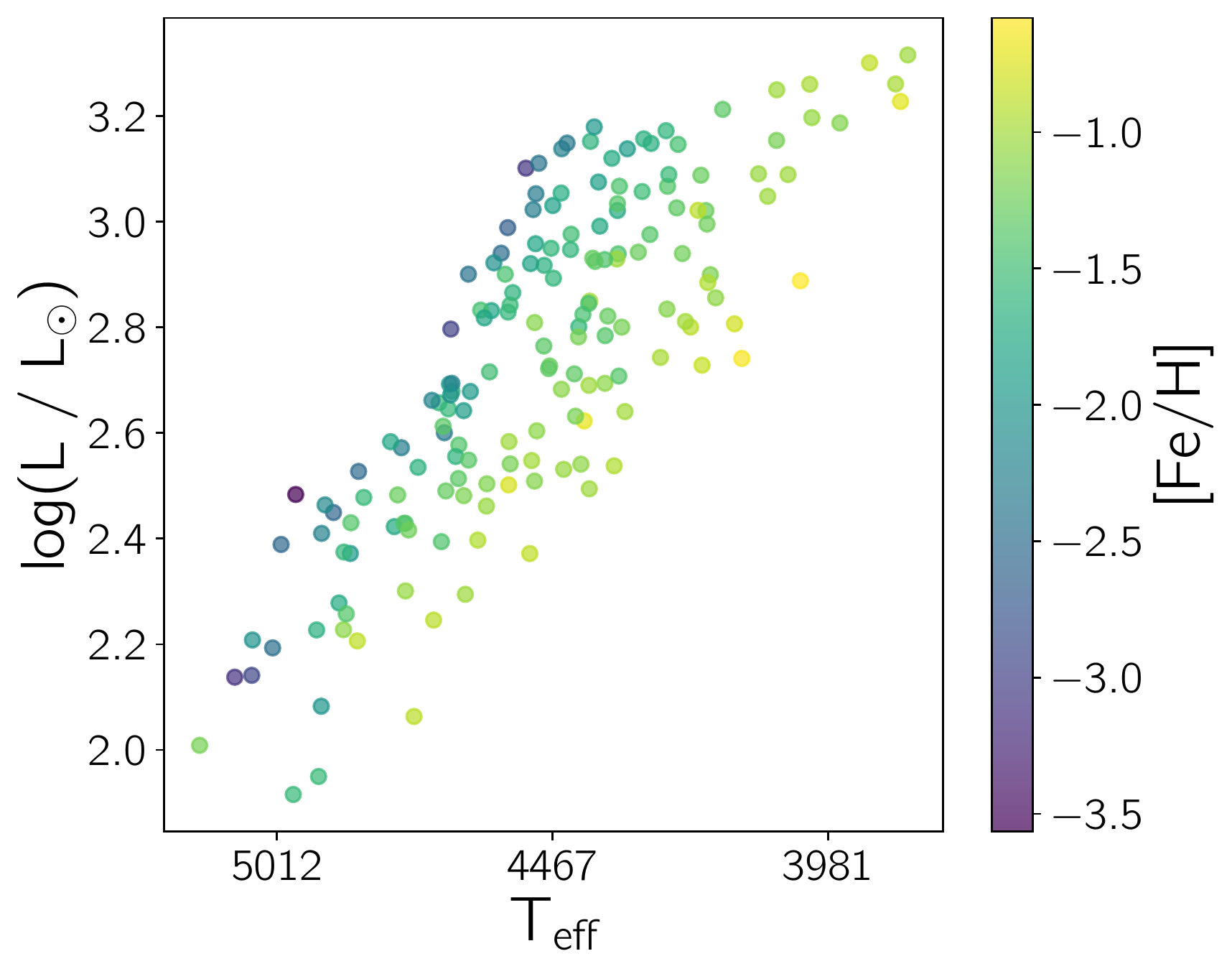}{0.31\textwidth}{(c)}
    \fig{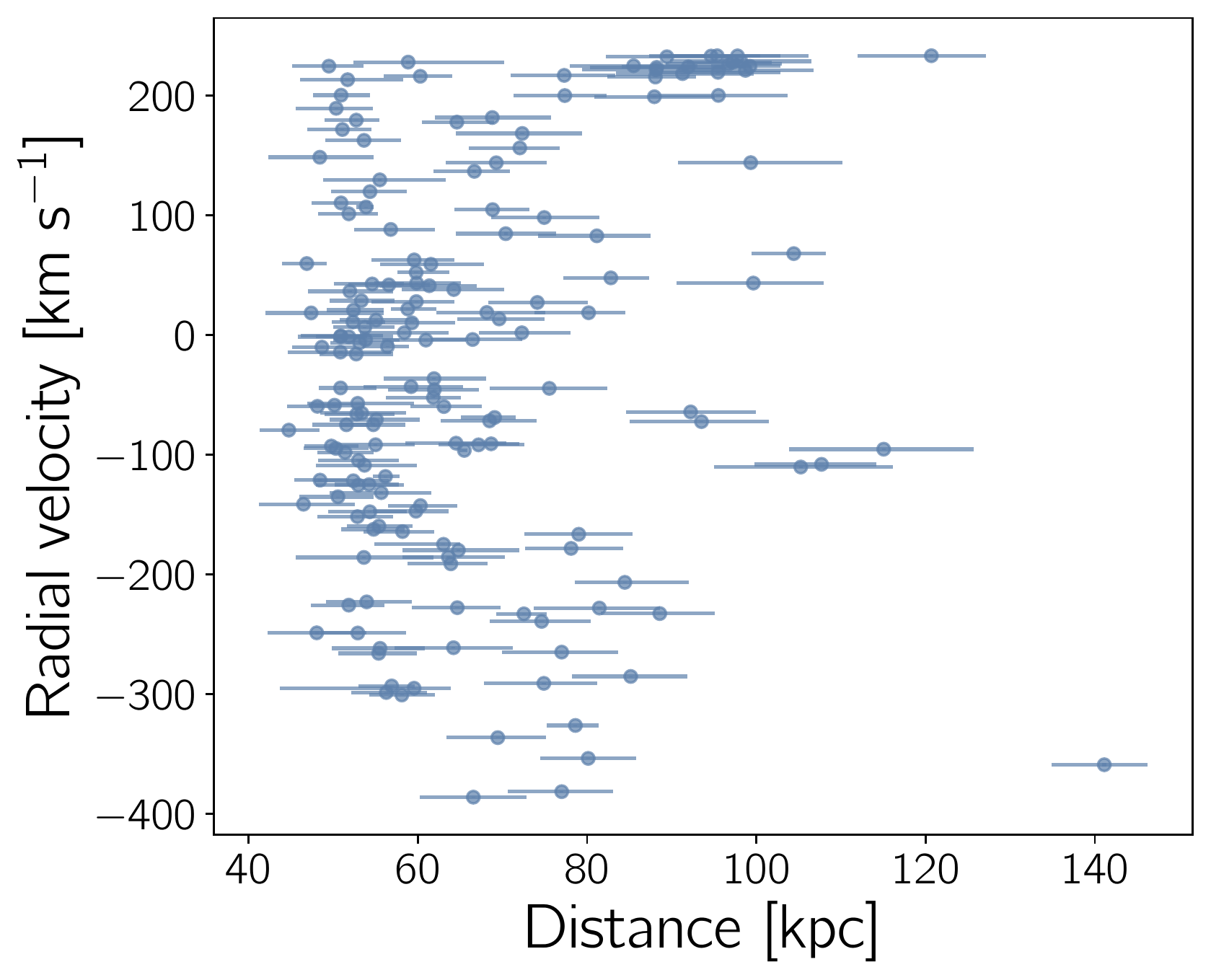}{0.31\textwidth}{(d)}
    \fig{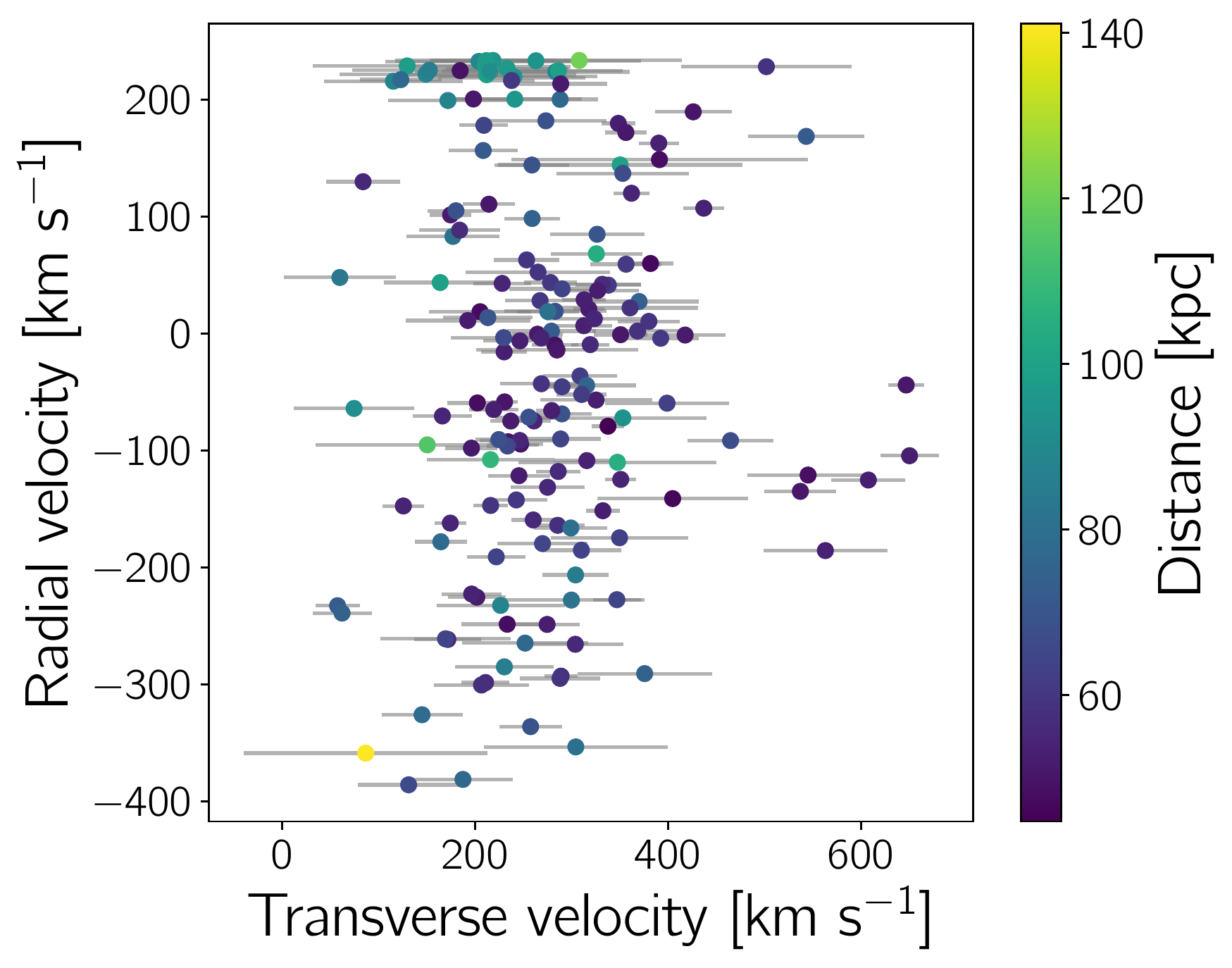}{0.31\textwidth}{(e)}
  }
  \caption{\textbf{(a)} Spatial distribution of stars in Galactic coordinates. The plot uses an Aitoff projection. \textbf{(b)} Spatial distribution of stars in Galactocentric coordinates, which shows where the stars would be located in a top-down view of the Galaxy. The height of the stars above or below the Galactic plane is indicated by the color of each point. The brown circle indicates the orbital radius of the Sun: \w{8.122}{kpc}. \textbf{(c)} Hertzsprung-Russell diagram of the selected sample. The x-axis shows the effective temperature and the y-axis shows the log luminosity in solar luminosities. The color of each point indicates its metallicity [Fe/H]\deleted{ according to the colorbar on the right}. The stars in the sample are K-giants. \textbf{(d)} Distances of stars plotted against their radial velocities. \textbf{(e)} Transverse velocities of stars plotted against their radial velocities.}
  \label{fig:params}
\end{figure*}

\added{Figure \ref{fig:params} shows various properties of the stars. Panel (a) shows the longitudes $\ell$ and latitudes $b$ of the sample on an Aitoff projected map. Panel (b) shows the spatial distribution of the stars in Galactocentric coordinates. The same plot contains a brown ring showing the radius of the Sun (\w{8.122}{kpc}); the Sun lies at the point where this ring intersects the x-axis.} The Hertzsprung-Russell diagram in \added{panel (c) of }Figure \ref{fig:params} shows that the selected stars have temperatures in the range of $4000-5000~{\rm K}$ and luminosities $100-2500$ times that of the Sun, suggesting that they are primarily K-giants. 

The stellar positions and proper motions \added{of the stars }are from \gaia\ EDR3, while radial velocities and spectrophotometric distances are from H3. The typical errors on the selected sample are \w{0.07}{mas} on right ascension (R.A.) and declination (Dec.), \w{0.1}{mas\,yr^{-1}} on proper motions ($\sim 10\%$), \w{0.3}{km\,s^{-1}} on radial velocities ($<1\%$), and \w{5}{kpc} on distance ($\sim 8\%$). \added{Panel (d) of Figure \ref{fig:params} shows the line-of-sight distances (i.e., distances from the Earth) plotted against the line-of-sight (radial) velocities, and panel (e) shows the transverse velocities (product of distance and total proper motion) plotted against the radial velocities. The errors for the transverse velocities only include the errors on the proper motions.}

  % \needspace{5\baselineskip}
  \section{Model Design}\label{sec:model} %%%%%%%%%%%%%%%%%%%%%%%%%%%%%%%%%%%

  \subsection{Multilevel Model and Distribution Function}\label{sec:model-design}

\begin{figure}[!htbp]
  \centering
  % \includesvg[width=1\linewidth]{gme.svg}
  \includegraphics[width=1\linewidth]{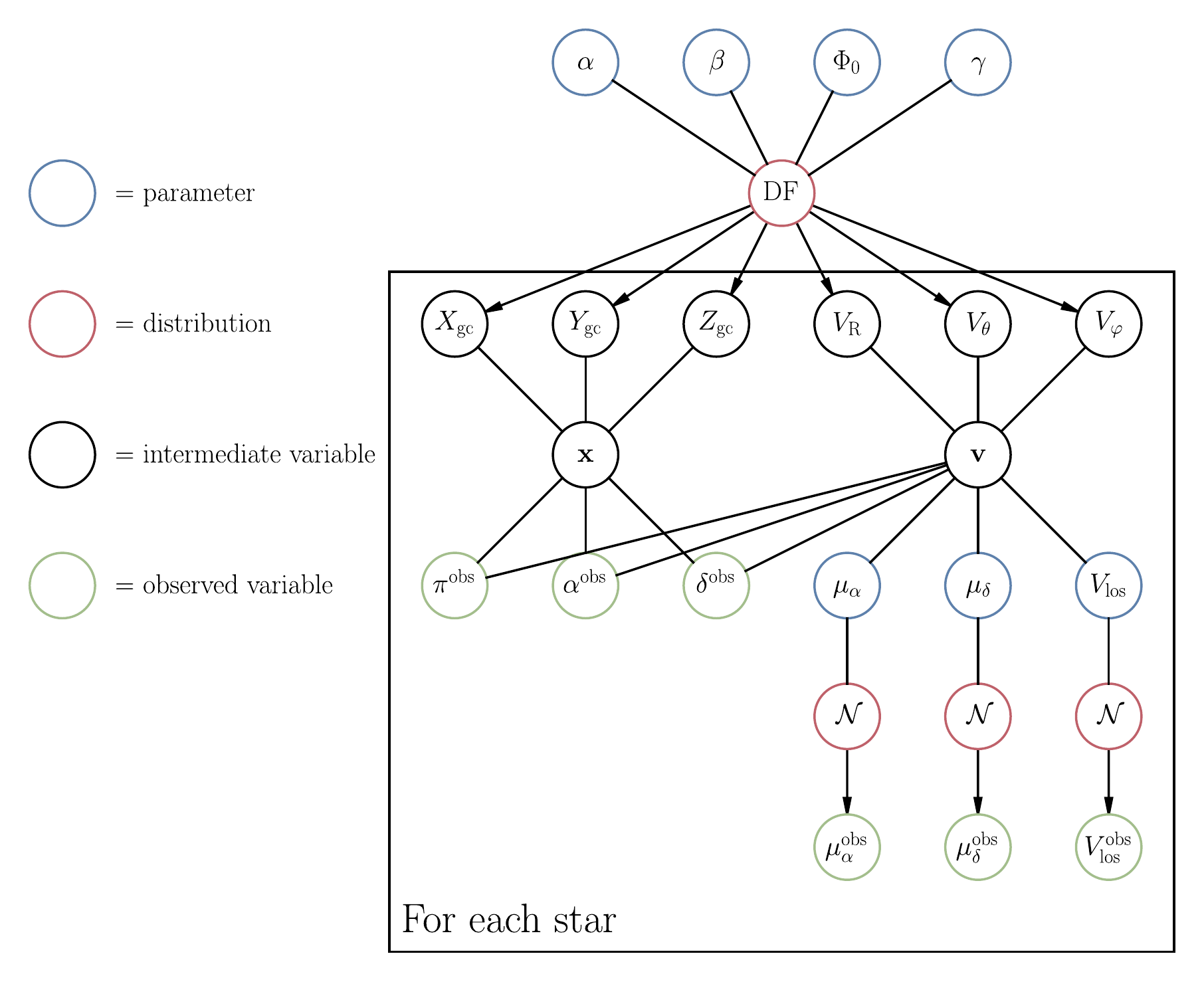}
  \caption{Graphical respresentation of \protect{\cite{Eadie2019}} model. Measured positions and distances are fixed, and measured velocities are assumed to be drawn from Gaussian distributions centered around true (latent) velocities.\label{fig:pgm-old}}
\end{figure}

  We first describe the model developed in \cite{Eadie2016} and \cite{Eadie2017}, which we will extend later on. This model has been used in two studies: \cite{Eadie2019}, which uses 32 globular clusters \citep{Vasiliev2019}, and \cite{Slizewski2021}, which uses 32 dwarf galaxies \citep{Fritz2018, Riley2019}. A graphical representation of the model is shown in Figure \ref{fig:pgm-old}. 

The Galaxy is modeled as a spherically symmetric system where the gravitational potential and the tracer population follow different power law slopes. The mass of the Galaxy can be characterized by four parameters:
\begin{itemize}
  \item $\Phi_0$: the scale factor for the gravitational potential,
  \item $\gamma$: the power law slope of the gravitational potential,
  \item $\alpha$: the power law slope of the tracer population, and
  \item $\beta$: the velocity anisotropy of the tracer population, which is assumed to be constant with Galactocentric radius.
\end{itemize}

The velocity anisotropy parameter is defined as
\begin{align}\label{eq:def_beta}
  \beta = 1 - \frac{\sigma_\theta^2 + \sigma_\phi^2}{2\sigma_r^2}, \quad \beta \in (-\infty, 1]
\end{align}
where $\sigma_r$, $\sigma_\theta$, and $\sigma_\phi$ are the dispersions (standard deviations) of the three velocity components in spherical coordinates (radial, azimuthal, polar). An isotropic system has $\beta = 0$, a radially dominated system has $\beta = 1$, and a rotationally dominated system has $\beta < 0$. 

The power law slope of the tracer population, $\alpha$, has the restriction that
\begin{align}\label{eq:restriction_alpha}
  \alpha > {\rm max}\{3,\, \beta\,(2 - \gamma) + \frac{\gamma}{2}\}
\end{align}
\citep{Evans1997}. The functional lower limit of 3 is a restriction on this particular analytical solution, and does not represent a physical restriction on the slope of the tracer population (i.e., in reality the slope may be less than 3). \added{We also note here that $\alpha$ does not represent the true physical distribution of stars in the halo. It is impacted by various biases in our sample, and furthermore, it is primarily a nuisance parameter that does not directly impact the estimated mass.}

Additionally, $\Phi_0$ is restricted so that the relative energy $\mathcal{E}$ is positive:
\begin{align}\label{eq:restriction_e}
  \mathcal{E} = -\frac{v^2}{2} + \frac{\Phi_0}{r^\gamma} > 0,
\end{align}
or equivalently,
\begin{align}\label{eq:restriction_phi0}
  \Phi_0 > \frac{v^2}{2}r^\gamma.
\end{align}

The distribution function for this model, which gives the probability of any particle being at some location in phase-space, is
\begin{align}\label{eq:def_df}
  \textstyle
    \mathcal{F}(\alpha, \beta, \gamma, \po) = \frac{ L^{-2\beta} \, \mathcal{E}^{\frac{\beta(\gamma - 2)}{\gamma} + \frac{\alpha}{\gamma} - \frac{3}{2}} \, \Gamma(\frac{\alpha}{\gamma} - \frac{2\beta}{\gamma} + 1)}{\sqrt{8 \pi^3 2^{-2\beta}} \, \Phi_0^{\frac{-2\beta}{\gamma} + \frac{\alpha}{\gamma}} \, \Gamma(\frac{\beta(\gamma - 2)}{\gamma} + \frac{\alpha}{\gamma} - \frac{1}{2}) \, \Gamma(1-\beta)},
\end{align}
where $L=r v_t$ is the total angular momentum, $r$ is the Galactocentric distance, and $v_t$ is the tangential velocity \citep{Binney1987, Evans1997} of a Galactic tracer. Note that the distribution function requires coordinates in the Galactocentric frame; see Section 2.3 of \cite{Eadie2017} or \cite{Johnson1987} for details on the heliocentric to Galactocentric transformation.

Positions in the heliocentric frame are assumed to be fixed. Velocities, on the other hand, are given a multilevel treatment: the observed velocity parameters are assumed to be drawn from a Gaussian distribution centered on latent velocity parameters. The measurement errors of the velocities are used as the standard deviations of the Gaussian. The latent velocities are then used in the coordinate transformation. 

With this model and $G \equiv 1$ units, the mass enclosed within any radius is given by 
\begin{align}\label{eq:mass_at_radius}
  \frac{\rm M(<r)}{10^{12}~{\rm M_\odot}} = 2.325 \times 10^{-3}\,\gamma\,\left(\frac{\po}{10^{4}~{\rm km^{2}\,s^{-2}}}\right)\,\left(\frac{r}{1~{\rm kpc}}\right)^{1 - \gamma}.
\end{align}
See Appendix \ref{appendix:mass} for more details. 

\needspace{3\baselineskip}
\subsection{New Model}\label{sec:new-model}

\begin{figure*}[!htbp]
  \centering
  % \includesvg[width=1\linewidth]{h3.svg}
  \includegraphics[width=1\linewidth]{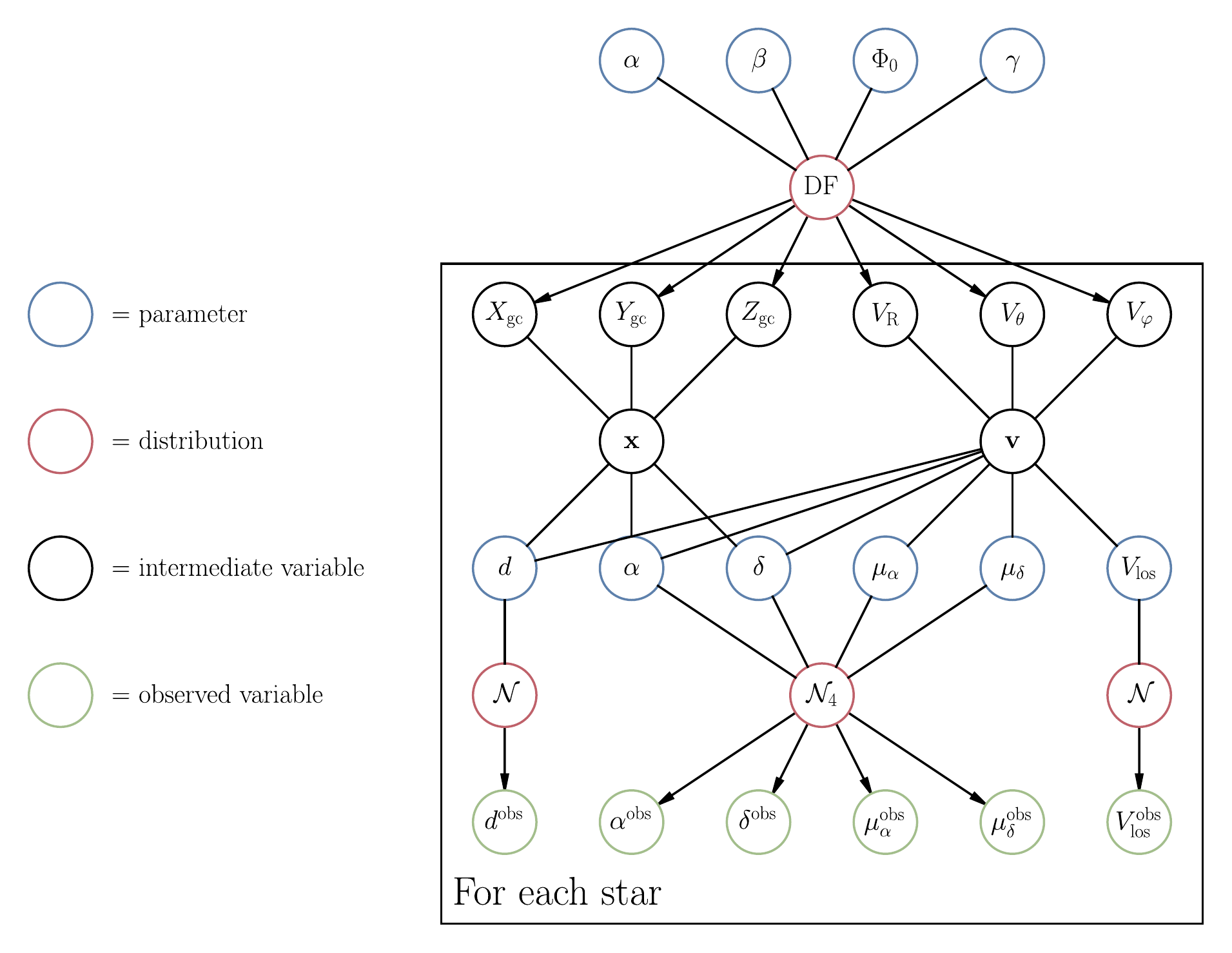}
  \caption{Graphical respresentation of the extended statistical model used for analysis of the H3 data. All six phase-space measurements are given a multilevel treatment; the observed variables are assumed to be drawn from Gaussian distributions centered around true (latent) phase-space parameters, with standard deviations equal to the measurement errors. For the positions and proper motions, in-source covariance information from \gaia\ is incorporated into the measurement error model. The heliocentric latent phase-space parameters are then transformed into the Galactocentric frame and used to estimate the four parameters of the distribution function.\label{fig:pgm-new}}
\end{figure*}

Although we use the model as described in Section \ref{sec:model-design} for comparison with \cite{Eadie2019} and \cite{Slizewski2021}, we make two major changes before applying it to the new dataset. The first is that the true positions (right ascension, declination, and distance) are treated as parameters, and the second is that we incorporate covariance information. A graphical representation of the new model is shown in Figure \ref{fig:pgm-new}.

Whereas the right ascension, declination, and heliocentric distances were previously treated as fixed, we now subject them to the same multilevel treatment as is given to the velocity measurements. We say that each measured distance (for example) is a sample that is drawn from a Gaussian distribution centered around some true distance---a latent variable---with standard deviation equal to the measurement uncertainty:
\begin{align}
  d^{obs}_i \sim \mathcal{N}(d_i, \sigma_{d,i}).
\end{align}

\added{We acknowledge here that because \gaia\ is able to obtain extremely precise measurements for the right ascension and declination, there is little benefit to incorporating the uncertainties for these. We include them for completeness, but in practice it would make more sense to keep these two parameters fixed for computational efficiency.}

Ideally we would use samples from the posteriors for the phase-space parameters estimated by \verb|MINESweeper|, but to first order the individual H3 posteriors can be well approximated by Gaussians. 

One benefit of this hierarchical treatment is that we obtain partially pooled estimates for the latent parameters \citep[see, e.g.,][]{Gelman2006}. These partially pooled estimates share information across stars, taking into account the measurement errors of each star, while still allowing for individual variation; this results in more reliable estimates of phase-space variables, and may be useful for follow-up studies. 

Because the distribution function is defined in terms of Galactocentric phase-space information, we require that the available heliocentric positions and velocities be transformed into Galactocentric positions and velocities each time the latent variables for each star are re-estimated (i.e., at each step in the Markov chain). This significantly increases the complexity of the model. The transformations that we use are consistent with those of Astropy v4.2.2 \citep{Astropy2013, Astropy2018}. The mathematics of the transformations are described in Appendix \ref{appendix:transformations}. 

A further improvement made to the model is that uncertainty covariances between the position and velocity parameters is incorporated into the observation process. \gaia\ provides within-source covariances between the positions, parallax, and proper motions of each source. By incorporating these into the observation process, our analysis takes advantage of all available information when performing inferences. Note here that since we do not directly use the \gaia\ parallaxes for distances, we also do not use the covariances provided for the parallax of each source. \added{The full expression for the posterior distribution is given in Appendix \ref{appendix:posterior}.}

With this structure, our model maximizes the available phase-space information. We are able to make mass estimates using any combination of sources missing positions, velocities, neither, or even both.
For example, for a source with no missing positions or proper motions, the measured positions and proper motions are assumed to be drawn from a 4-dimensional multivariate normal distribution with the appropriate covariances between all four parameters. However, if a source is missing position measurements, we can still incorporate the covariances between the measured proper motions. In the case of the H3 data, we have complete 6D information for all stars.

  % \needspace{5\baselineskip}
  \section{Inference with HMC}\label{sec:inference} %%%%%%%%%%%%%%%%%%%%%%%%%%%%%%%%%%%

  \subsection{Introduction to Stan}\label{sec:inference-stan} %%%%%%%%%%%%%%%%%%%%%%%%%%%%%%%%%%%

Stan is an open-source probabilistic programming language for Bayesian inference, with interfaces in both Python and R (among others). It implements several inference algorithms, with the most notable being the No-U-Turn sampler (NUTS), an extension of Hamiltonian Monte Carlo (HMC). The advantages of HMC are that it is fast, scalable, and easy to troubleshoot. The Stan code for all models used in this paper, as well as examples for calling the models from both Python and R, can be found at \url{https://github.com/al-jshen/gmestan-examples}.

\needspace{2\baselineskip}
\subsection{Hamiltonian Monte Carlo}\label{sec:inference-hmc} %%%%%%%%%%%%%%%%%%%%%%%%%%%%%%%%%%%

Metropolis-Hastings algorithms \citep{Metropolis1953, Hastings1970}, a class of Markov chain Monte Carlo (MCMC) algorithms, operate by making proposals to new positions in parameter space and checking whether the posterior at the proposed position is favorable relative to the current position. 
If the proposal method is poorly chosen, these algorithms will tend to waste computing power (e.g., get stuck in the same location or inefficiently explore parameter space). 
Traditional MCMC algorithms such as random-walk Metropolis (RWM) and Gibbs samplers \citep{Geman1984, Casella1992} tend to use Gaussian proposal distributions, leading to random-walk behavior. 
Tuning the proposal distribution by hand to achieve efficient sampling is also time consuming and challenging.

\added{\texttt{emcee} \citep{Foreman-Mackey2013}, a popular package for posterior sampling, also
struggles in high dimensional problems. \cite{Huijser2017} show that even for a
simple correlated Gaussian, \texttt{emcee} is unable to recover the expected mean
and variance in just 100 dimensions even after the MCMC is run for 200,000
iterations. \cite{Speagle2019} also show that the stretch move technique used
by \texttt{emcee} makes exponentially fewer good proposals as the dimensionality
of the problem increases. The explanation for why this technique performs
poorly in high dimensions is that the stretch move proposal is an
interpolation/extrapolation between one walker and another. Consider, for
example, a high dimensional Gaussian, where the typical set (see paragraphs
below) lies in a thin shell surrounding the mode. If we take two points on this
shell and then interpolate or extrapolate between them, a point on this line 
is unlikely to lie in the typical set unless it is close to one of the original
points. Thus the algorithm essentially devolves into a random walk.}

HMC is a gradient-based MCMC algorithm that avoids random-walk behavior by making better proposals using Hamiltonian dynamics \citep{Duane1987, Neal2011}. A simulated particle is placed on a frictionless surface (the negative log-posterior), and its motion on that surface is simulated with auxiliary momentum parameters. A random direction and energy are chosen and assigned to the particle. After some time, the particle's new location in parameter space is taken as a sample \citep{Rethinking}. By using the gradient of the log-posterior to inform the movement of the particle, HMC avoids staying in the same position for long periods due to bad proposals, and ensures that subsequent proposals are distant in parameter space \citep{Neal2011, Betancourt2015, Betancourt2017a}. To solve the Hamiltonian equations, a numerical integrator is used. In practice, the leapfrog integrator is chosen, both because it is time-reversible \footnote{This will be particlarly important later in NUTS.} and because it is symplectic, meaning that it preserves energy over long integration times.

HMC provides numerous benefits over RWM\added{ and ensemble methods}, particularly in high-dimensional problems.
Probability mass---the product of probability density and probability volume---tends to be concentrated in the ``typical set'', a surface in between the mode (region with the highest probability density) and the tails (region with high probability volume). As the dimensionality of the problem increases, more and more of the probability volume exists in the tails of the distribution, and consequently the typical set becomes smaller (i.e., the region of parameter space where there is overlap between the probability density and probability volume becomes smaller). Whereas RWM will tend to get stuck in one region of parameter space, either due to large step sizes which result in constant rejection or small step sizes which lead to inefficient exploration, HMC is able to exploit the geometry of the typical set to efficiently explore parameter space. We refer the reader to \cite{Betancourt2017a}, \cite{Carpenter2017a}, and \cite{Speagle2019} for excellent explanations of typical sets and the difficulties of sampling in high dimensions.

There are two main drawbacks to HMC. The first is that, per sample, it is much more computationally expensive than RWM. HMC requires computation of the gradient, which is expensive---although automatic differentation makes this tractable \footnote{Automatic differentiation is free of the rounding and truncation errors introduced by numerical differentiation and the complexity of expressions (so called ``expression swell'') of symbolic differentiation. Automatic differentiation is essentially an aggressive application of the chain rule to the elementary operations (e.g., addition, multiplication, exponentiation, logarithm, sines, cosines) which make up any computation \citep{Wengert1964, Griewank2008, Carpenter2015, Baydin2017, pytorch}.}. This disadvantage is partially offset by the lack of need for ``thinning'', since successive samples have low autocorrelation. In other words, the number of samples to achieve the same effective, de-correlated sample size is typically far lower, and traditional recommendations to run chains for tens or hundreds of thousands of steps is no longer necessary.

The second drawback of HMC is that the gradient, through the geometry of the target distribution, depends on the specific parameterization of the problem. This means that the step size, the Euclidean metric (which accounts for linear correlations in the posterior), and the number of steps in each HMC iteration need to be tuned manually in order for HMC to perform well (see Section 15.2 in the Stan Reference Manual for more details; \citealt{stanmanual}). If the step size is too small, then the target density may not be explored effectively. On the other hand, if the step size is too large, then regions of the posterior where the probability density is highly concentrated in a small volume may not be well resolved. Furthermore, choosing too small a number of steps for each iteration means that subsequent samples may be very close together, leading to degeneration to random-walk behavior. Too many steps and the trajectory of the simulated particle may loop back to a location it has already explored, meaning that expensive gradients are needlessly calculated. In the worst case of a poorly tuned HMC, the particle may repeatedly loop back in a way that subsequent samples are very close together. This is the ``U-Turn problem.''

\subsection{The No-U-Turn Sampler}\label{sec:inference-nuts} %%%%%%%%%%%%%%%%%%%%%%%%%%%%%%%%%%%

NUTS is an adaptive extension of HMC that removes the need to manually select the HMC parameters. During a warmup phase, Stan uses a modified version of the dual averaging algorithm by \cite{Nesterov2009} to adaptively tune the step size. The linear correlations in the posterior are also estimated during warmup. To determine the number of steps to take from an initial position, NUTS chooses a standard normal momentum vector, and then integrates both forward and backward in time, which is possible because the integrator was chosen to be time-reversible. The integrator takes one step either forward or backward, then two steps either forward or backward, then four steps either forward or backward, and so on. NUTS builds a balanced binary tree while doing so, with each of these iterations adding a balanced subtree and increasing the total tree depth by one. When two nodes in any subtree form a U-turn---indicated by an angle greater than $90^\circ$ between corresponding momentum vectors of the two nodes---the algorithm stops and carefully takes a sample in a way that maintains detailed balance \citep[see Figures 1 and 2 in][]{Hoffman2014, stanmanual}.

What all this means is that (1) the end user only needs to worry about the design of their model and not about computational inefficiencies, and (2) inference for large models is computationally tractable. The model we use for the H3 dataset has over a thousand parameters.

\needspace{5\baselineskip}
\section{Calibration with Mock Data}\label{sec:discussion-mocks}

Before analyzing the real H3 data, we first use H3-mocked catalogs (R. Naidu, private comm.) based on the Auriga simulations \citep{Grand2019} as a test to see how well our method can recover masses. We use the simulated galaxies Au6, Au23, Au24, and Au27, which include the H3 selection function. Each of the galaxies also has measurement errors in the derived quantities that are comparable to what is expected in the H3 data. For these catalogs, we select stars based on the same criteria used for the real data (see Section \ref{sec:data}).

One benefit of Bayesian analysis is the ability to incorporate prior beliefs---or prior results---into our analysis. To carry over information from previous studies, we use the posteriors for $\po$ and $\gamma$ from \deleted{the dwarf galaxy study}\added{a previous study} (see Section \ref{appendix:stancheck}) as hyperpriors for our new analysis. These parameters describe the Galaxy's potential, and should be independent of the tracer population.
Because Stan requires that priors be specified with analytical formulae, we fit a multivariate Gaussian to the posterior densities of $\gamma$, and $\po$---which retains the correlation compared to using two univariate Gaussians fitted to the marginal posteriors---and use that approximation rather than the full posterior. Note here that this really is an approximation, and nuances in the posterior are not captured.

On the other hand, we set hyperpriors for $\alpha$ and $\beta$ separately based on information from other studies rather than using the posteriors from the dwarf galaxy study. This is because these two parameters are dependent on the tracer population, and dwarf galaxies are not halo stars. For $\alpha$, we apply a prior of $\alpha \sim \mathcal{N}(4.0, 0.1)$, broadly following \cite{Deason2021} (who set $\alpha = 4$). Based on recent estimates for the anisotropy of the Galaxy from \cite{Bird2019}, we choose a prior of $\beta \sim \mathcal{N}(0.3, 0.1)$, favoring radial orbits. We note that these two parameters are primarily nuisance parameters and they have their limitations (some of which we will discuss in Section \ref{sec:discussion-ppc}); we are not currently too interested in them, but they must be included as per Eq. \ref{eq:def_df}. The hyperpriors used for the analysis are listed in Table \ref{table:hyperpriors-h3}.

\begin{deluxetable}{ccc}[!ht]
  \tabletypesize{\scriptsize}
  \tablecaption{Hyperprior distributions used for the H3 halo star analysis. \label{table:hyperpriors-h3}}
  \tablehead{
    \colhead{Model Parameter} & \colhead{Distribution} & \colhead{Distribution parameters}
  }
  \startdata
  $\alpha$ & Normal & $\mu = 4$, $\sigma=0.1$ \\
  $\beta$  & Normal & $\mu = 0.3$, $\sigma=0.1$ \\
  $\begin{bmatrix} \po \\ \gamma \end{bmatrix}$ & Multivariate Normal & $\mu = \begin{bmatrix} 64.63 \\ 0.44 \end{bmatrix}$, $\Sigma = \begin{bmatrix} 47.61 & 0.165 \\ 0.165 & 0.0021 \end{bmatrix}$ \\
  \enddata
  % \tablecomments{}
\end{deluxetable}

For each simulated galaxy, we perform split-sky tests in addition to analyses with the selected sample. Briefly, these split-sky tests are intended to test whether our mass estimates are affected by spatially coherent substructure, and are performed by removing stars in the sample that belong to a particular quadrant of the sky, based on right ascension. We point the reader to Section \ref{sec:discussion-splitsky}, where we perform split-sky tests on the real data, for more details.

Given that the errors on the positions are extremely small (on the order of $10^{-8}~{\rm deg}$), we artificially inflate them in the 4D covariance matrix when sampling positions and proper motions to obtain greater numerical stability. We also performed a run where the positions were treated as fixed, and found no appreciable difference in the resulting posterior distributions.

The results of all the analyses on the mock Auriga galaxies are shown in Table \ref{table:mocks} in Appendix \ref{appendix:mocks}. With the exception of one run in Au6 where stars with R.A.\footnote{The right ascension for the mock data range from $-180^\circ$ to $180^\circ$, compared to the H3 data where the right ascension varies from $0^\circ$ to $360^\circ$.} from $-180^\circ$ to $-90^\circ$ were removed, all the mass estimates within each simulated galaxy are consistent within \deleted{$1\,\sigma$}\added{the 68\% credible interval} (and for that run, the results are consistent with the other results within $2\,\sigma$). We thus conclude that substructure, to the degree present in this simulated galaxy, does not affect our mass estimates.

Raw snapshot data are available for Au6.\footnote{\url{https://wwwmpa.mpa-garching.mpg.de/auriga/}} We use the raw snapshot data to construct a ``true'' mass curve by binning dark matter particles, star particles, and gas cells into $1~{\rm kpc}$ bins based on their distance from the center of the simulated galaxy. The total mass of all material in each bin is summed up to give the mass at that radius. The total mass enclosed within some radius $R$ is simply the sum of the masses of all the bins with radius less or equal to $R$.

\begin{figure*}[htbp!]
  \centering
  \gridline{%
    \fig{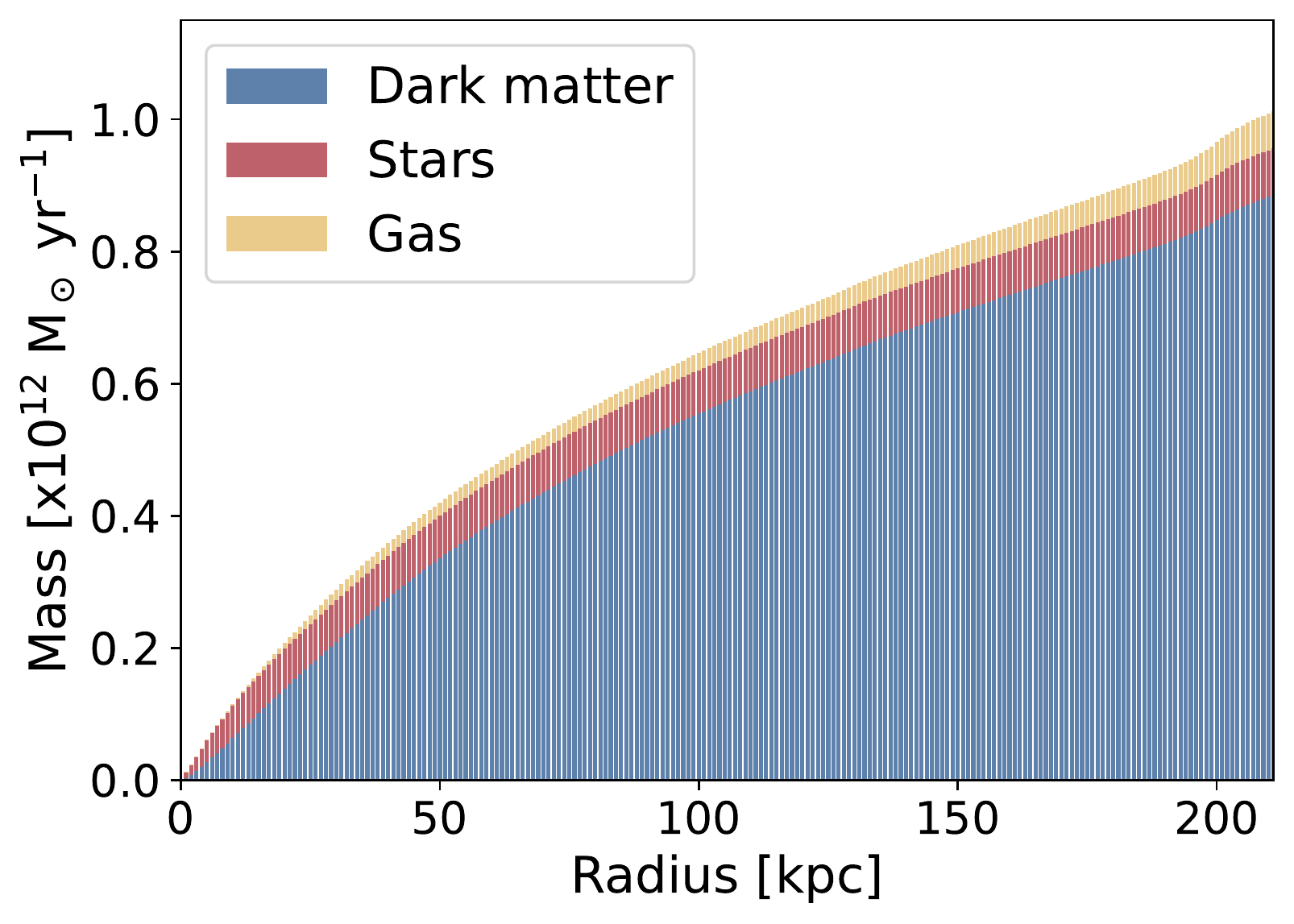}{0.33\textwidth}{(a)}
    \fig{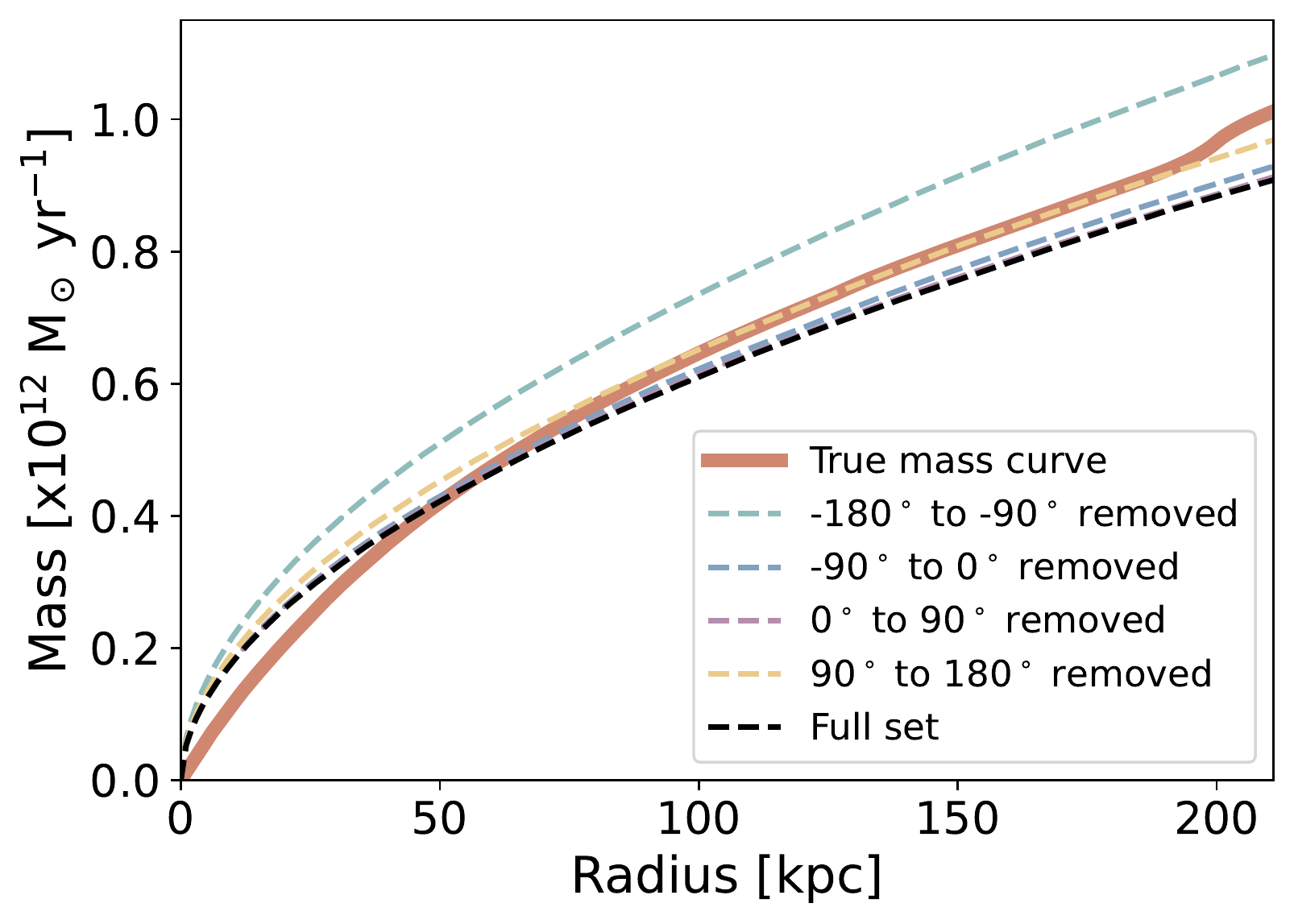}{0.33\textwidth}{(b)}
    \fig{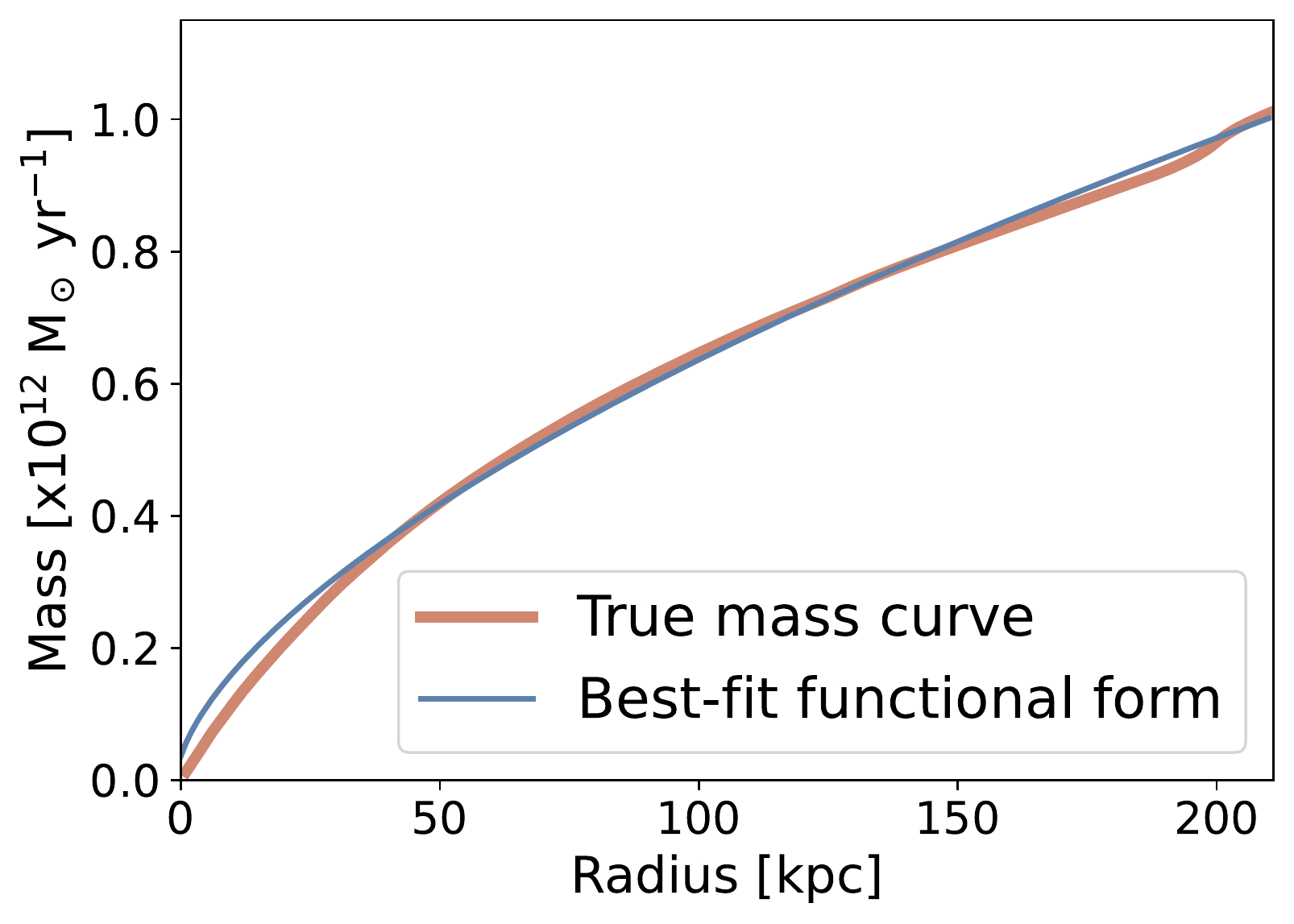}{0.33\textwidth}{(c)}
  }
  \caption{(a) True cumulative mass profile for Au6, broken down by particle type. The plot also contains the mass of a black hole. (b) True cumulative mass profile for Au6 (with all particle types included) plotted together with the estimated mass profiles from the four split-sky runs and a run with the full H3 mocked sample. See also Table \ref{table:mocks}. (c) True cumulative mass profile plotted along with the best-fitting line using Equation \ref{eq:mass_at_radius}. The estimated masses are in good agreement in the range of the data (roughly $50-100~{\rm kpc}$), and slightly underestimated when extrapolating out to larger radii.}
  \label{fig:snapshot}
\end{figure*}

Panel (a) in Figure \ref{fig:snapshot} shows this true cumulative mass profile broken down by particle type. The mass is dominated by dark matter at all radii. The mass of a black hole, although not labeled, is included in this plot; this represents a mass of $9.47\times 10^{7}~{\rm M_\odot}$ in the \w{1}{kpc} bin. In panel (b) the true mass profile of Au6 is plotted in red, together with the different estimated mass curves from the four split-sky runs and the estimate from the full set which are plotted with dashed lines. This panel shows that the masses from the model are overestimated at small radii (within $\sim 40~{\rm kpc}$) but slightly underestimated at large radii, with the exception of one split-sky run---the run where stars with RA from $-180^\circ$ to $-90^\circ$ are removed.

The full sample has stars going out to a Galactocentric radius of \w{117}{kpc}, with 95\% of the stars within \w{100}{kpc}. The estimated mass within \w{100}{kpc} using the full sample is $\rm M(<100\,kpc) = 0.61_{-0.06}^{+0.07} \times 10^{12} \; M_\odot$ in good agreement with the true mass of $\rm M(<100\,kpc) = 0.65 \times 10^{12} \; M_\odot$. Extrapolating out, $\rm M_{200}$ from the full sample is roughly 10\% lower than the true mass at a comparable radius ($\sim 203~{\rm kpc}$), at $\rm M_{200} = 0.89_{-0.12}^{+0.14} \times 10^{12} \; M_\odot$ and $\rm M_{200} = 0.98 \times 10^{12} \; M_\odot$ respectively; the estimates are consistent within \deleted{$1\,\sigma$}\added{the 68\% credible intervals}. This 10\% variation is also similar to the variation in the $\rm M_{200}$ between the different split-sky runs for this galaxy. Note that this estimated $\rm M_{200}$ requires a higher degree of extrapolation than the real data, as the simulated data do not go out as far (\w{117}{kpc} compared to \w{142}{kpc}).

Given that we only have the true mass profile for Au6, it may also be possible that our model only performs well in recovering the mass for this particular galaxy, and the model would perform less favorably on other galaxies. Further investigations into simulated data where the true mass distributions are known would be a worthwhile pursuit to determine (1) how well the model performs on average, and (2) what may influence whether the model underestimates or overestimates masses (e.g., a study similar to \citealt{Eadie2018}).

As another check, we directly fit the functional form of our mass profile equation (see Eq. \ref{eq:mass_at_radius}) to the true mass curve of Au6 using the Levenberg-Marquardt algorithm \citep{Levenberg1944,numericalrecipes}. This represents the best possible case given our functional form. Both the true mass curve and the best-fit line are shown in the rightmost panel of Figure \ref{fig:snapshot}. The best-fit functional form overestimates the mass at small radii, which is the same behavior that is observed in Panel (b) of the same figure. However, it is generally in good agreement with the true mass curve beyond $\sim 50~{\rm kpc}$. This suggests that the functional form we are using is capable of obtaining a relatively accurate cumulative mass profile at larger radii, and the discrepancies between the estimated and true mass curves that we observe are likely caused by the data or limitations of the model (see Section \ref{sec:discussion-ppc}).

  % \needspace{5\baselineskip}
  \section{Analysis of H3 Data}\label{sec:analysis} %%%%%%%%%%%%%%%%%%%%%%%%%%%%%%

  We now turn our attention to the primary analysis in our work. We analyze the real H3 halo star data described in Section \ref{sec:data}. We use the same prior distributions given in Table \ref{table:hyperpriors-h3}.

\begin{figure*}[htb!]
  \centering
  \includegraphics[width=0.97\linewidth]{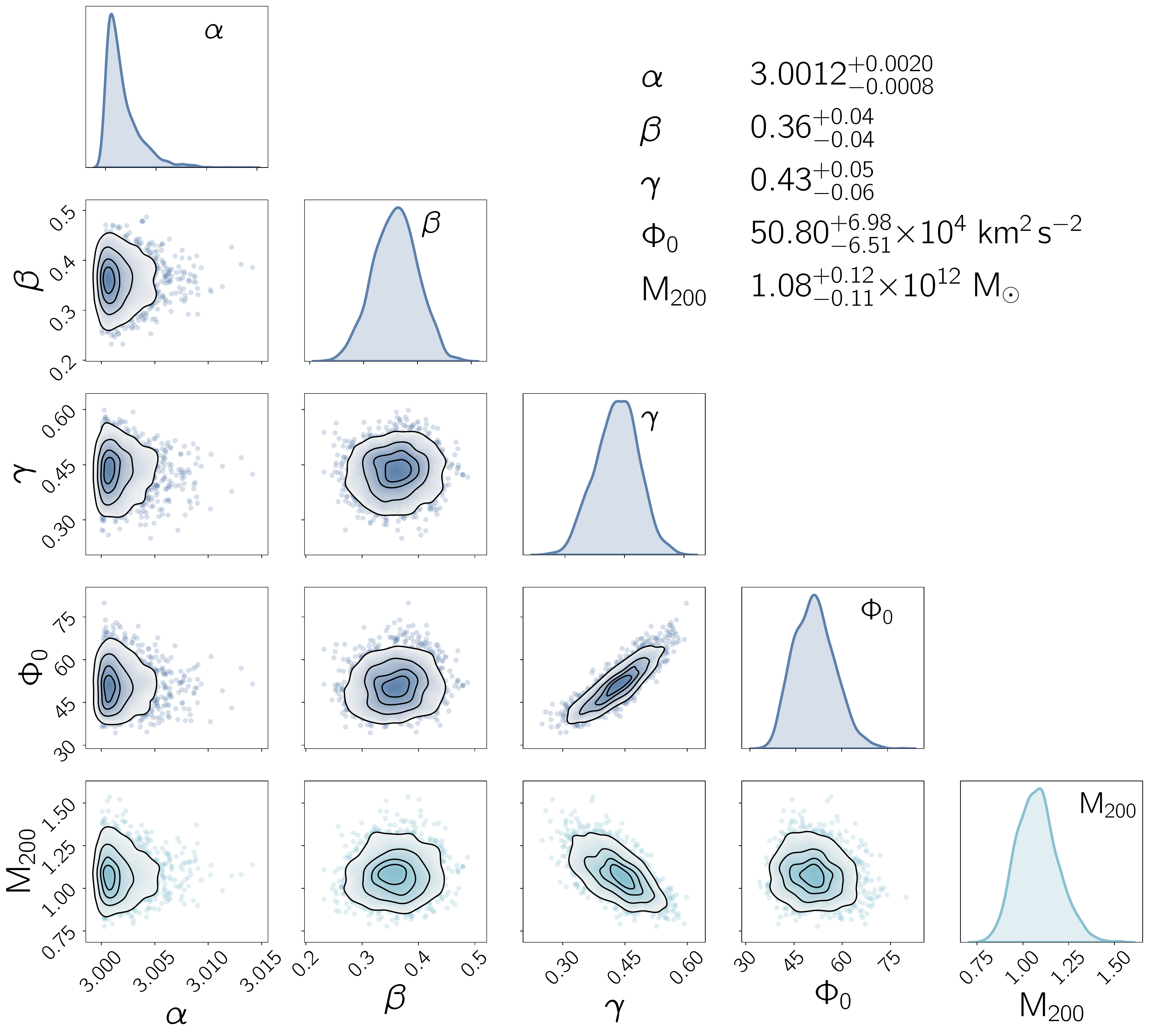}
  \caption{Pairs plot for the H3 data showing the marginal posterior densities for each model parameter and the joint distributions for each pair of parameters, given the H3 data and our prior choices. The bottom row gives the estimated total mass of the Galaxy, and is shown in a different color to indicate that it is a derived parameter, rather than one that is estimated directly in the model. Along the diagonals are kernel density estimates of the marginal posterior distributions for each estimated parameter. In the lower diagonal panels, the shading represents the 2D kernel density estimates of the joint posterior distributions for each pair of parameters. The contours, from the outermost to the innermost, contain 11.8\%, 39.3\%, 67.5\%, and 86.5\% of the probability volume (these contours correspond to 0.5, 1.0, 1.5, and 2.0 $\sigma$ in the case of a 2D Gaussian; see \protect \url{https://corner.readthedocs.io/en/latest/pages/sigmas.html}). A scatterplot for data points outside of the outermost contour is also shown. The values quoted in the upper right are the median estimates, and the uncertainties represent the 16th and 84th percentiles. The units for $\po$ are $10^{4}~{\rm km^{2}\,s^{-2}}$ and the units for $\rm M_{200}$ are $10^{12}~{\rm M_\odot}$; the other three parameters are unitless.}
  \label{fig:h3_pairs}
\end{figure*}

The estimated posteriors for our study using the H3 data are shown in Figure \ref{fig:h3_pairs}. The diagonals show the marginal posterior densities for $\alpha$, $\beta$, $\gamma$, and $\po$, and the panels below the diagonals are scatterplots for each pair of parameters, with kernel density estimates overplotted.

The model, with 1012 free parameters (6 for each tracer and 4 for the distribution function), takes roughly 10 minutes to run on an AMD Ryzen 5 3600 CPU using the default sampling configuration in CmdStanPy (1000 post-warmup draws per chain). This kind of computation would have been prohibitively expensive to run with GME, considering that it takes hours to run a model with an order of magnitude fewer parameters. The posterior chains were checked with diagnostics to ensure that divergences, if any, are not concentrated in parameter space (which would bias inference), and that the $\hat{R}$ and effective sample size are satisfactory.

The marginal posterior distribution for $\alpha$ is highly skewed, with probability mass piling up at 3, which is the statistical lower bound allowed by our model. We note that this is somewhat of a nuisance parameter that has no direct impact on the mass profile. This skewness in the posterior is likely caused by model mismatch, which we discuss further in Section \ref{sec:discussion-ppc}.

We also performed prior sensitivity analyses, and found that our results are robust to changes in the priors. We tested priors of, for example, $\beta \sim \mathcal{N}(0, 0.1)$, $\beta \sim \mathcal{N}(0.3, 0.05)$, $\alpha \sim \mathcal{N}(3.5, 0.05)$, \deleted{and }$\alpha \sim \mathcal{N}(4, 0.05)$, \added{and $\alpha \sim \mathcal{N}(4.4, 0.1)$, }and each time the parameter\deleted{s} estimates were not meaningfully different from what is reported here.

The estimate for the velocity anisotropy parameter is $\beta = 0.36_{-0.04}^{+0.04}$, which indicates slightly radial orbits in the outer halo. This value does not seem to be very sensitive to the selected prior based on tests with several priors centered at different values and with different widths than our default. Our estimate for $\beta$ is in very good agreement with \cite{Kafle2014a}, who find that $\beta = 0.4_{-0.2}^{+0.2}$ using 5140 giants in the halo from the ninth data release of the Sloan Digital Sky Survey (SDSS, \citealt{Ahn2012}). It also agrees well with a more recent estimate from \cite{Bird2019}, who use 7664 K giants from the fifth data release of the Large Sky Area Multi-Object Fiber Spectroscopic Telescope (LAMOST, \citealt{Cui2012}) to model $\beta$ as a function of radius, and find that $\beta = 0.3$ at $\sim 100~{\rm kpc}$.

We find an enclosed mass within \w{100}{kpc} of ${\rm M(<100 \; kpc) = 0.69_{-0.04}^{+0.05}\times 10^{12} \; M_\odot}$. Note that although our formulation of the mass in terms of $\gamma$ and $\po$ allows us to estimate the mass at arbitrary radii, the data used to estimate these parameters only go out to $140~{\rm kpc}$, with 93\% of the data at ${\rm R_{gal} < 100 \; kpc}$. Thus, any estimates of the mass at larger radii are extrapolations. With this in mind, we report a median estimate for the total mass of the Milky Way of ${\rm M_{200} = M(<216.2_{-7.5}^{+7.5} \; kpc) = 1.08_{-0.11}^{+0.12}\times 10^{12} \; M_\odot}$.

\added{We also fit a Navarro-Frenk-White (NFW; \citealt{Navarro1997}) profile to our estimated mass profile from \w{50}{kpc} to \w{100}{kpc} (the range of the bulk of our data) and extrapolate for $\rm M_{200}$ based on this NFW profile. We find a virial mass of $\rm M_{200} = \rm M(<207.0_{-6.0}^{+6.0}\;kpc) = 0.94_{-0.08}^{+0.08}\times 10^{12} \; M_\odot$---this is roughly $12\%$ lower than the extrapolation based on Eq. \ref{eq:mass_at_radius}, but within the 68\% credible intervals.}

For ease of comparison, estimated masses at \w{10}{kpc} increments out to \w{250}{kpc} \added{(extrapolations are made with Eq. \ref{eq:mass_at_radius}) }are provided in Table \ref{table:masses} in Appendix \ref{appendix:masses}.

\added{As noted before, despite the furthest star in our sample being at \w{142}{kpc}, the majority of our data do not extend out that far. We are thus interested in seeing how informative those distant stars are in our fit; if they were to be removed, would our parameter estimates differ significantly? On the other hand, what estimates would we obtain if we only used distant stars? We perform a run using only stars within a Galactocentric radius of \w{80}{kpc}, and a run using only stars outside of \w{80}{kpc}. For the first test, we find that the estimates for all parameters to be, within the uncertainties, the same as the estimates using the full dataset. For the $r > 80~{\rm kpc}$ run, the estimate for $\po$ is larger than the estimate using the full dataset, and as a result the estimate for the virial mass is higher, at $\rm M_{200} = 1.45_{-0.17}^{+0.20} \times 10^{12}~{\rm M_\odot}$. Given that the mass estimate using the full dataset is essentially identical to the estimate from the $r < 80~{\rm kpc}$ analysis, it seems that the distant stars do not have a strong influence on the fit. A possible explanation for this is that the errors for the measurements for the distant stars (in particular the distance errors) are larger, and thus the hierarchical model applies more shrinkage to these stars.}

  \needspace{5\baselineskip}
  \section{Discussion}\label{sec:discussion} %%%%%%%%%%%%%%%%%%%%%%%%%%%%%%%%%%

  \needspace{2\baselineskip}
\subsection{Results in context}\label{sec:discussion-comparison}

\begin{figure*}[htb!]
  \centering
  \includegraphics[width=0.85\linewidth]{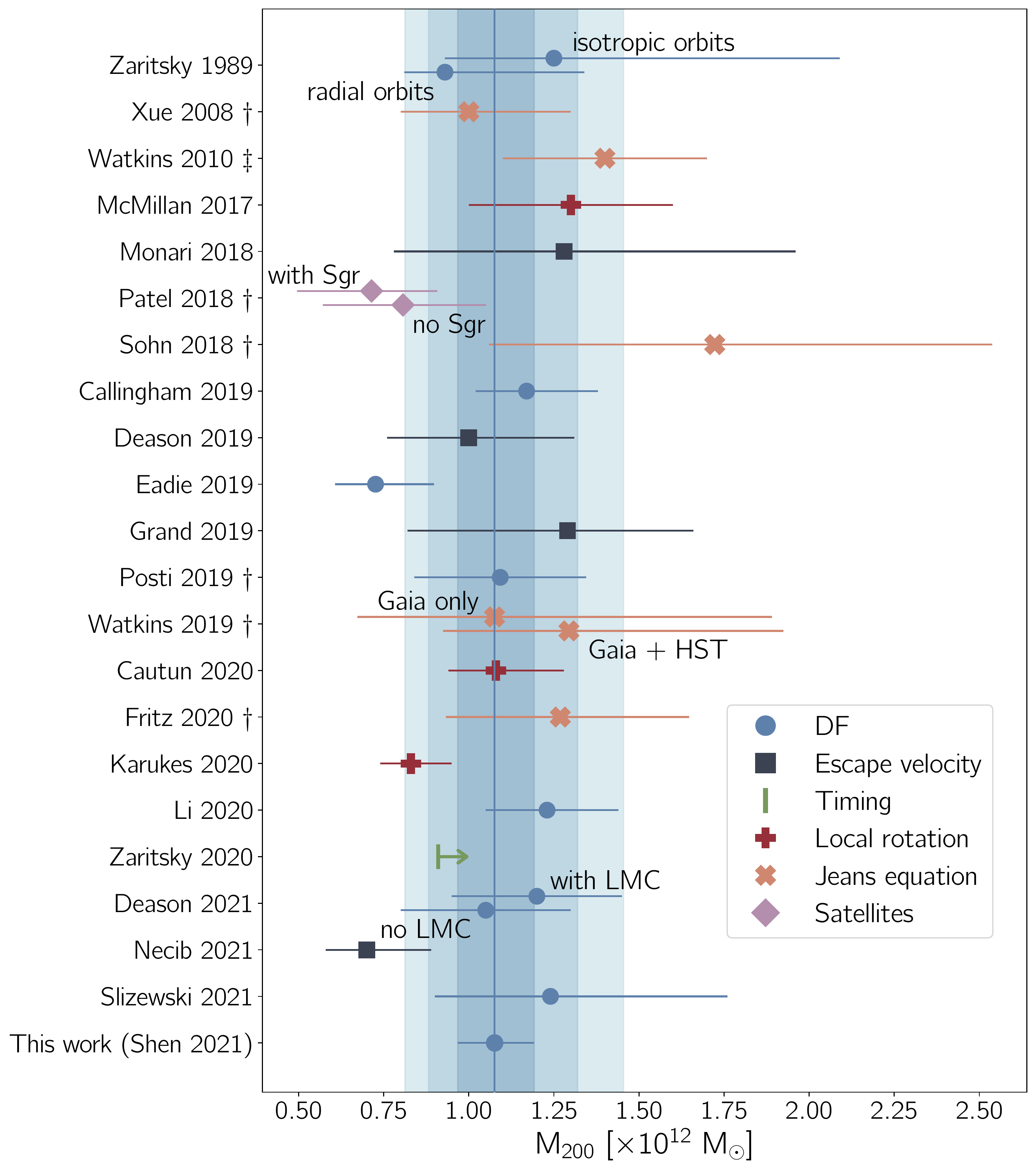}
  \caption{A compilation of recent literature measurements for ${\rm M_{200}}$ in order of year of publication, along with our estimated mass.
    The solid blue vertical line is our median ${\rm M_{200}}$, and the three shaded bands are the 68\%, 95\%, and 99.7\% credible intervals.
    The various points represent mean/median $\rm M_{200}$ measurements and reported $68\%$ errors from the literature\deleted{ (except for \citealt{Eadie2019}, marked with a $\ast$, which reports errors from the $50\%$ credible interval)}, where masses from studies which report ${\rm M_{vir}}$ with overdensities other than $\Delta_{\rm vir} = 200$ (marked with $\dagger$) have been converted to ${\rm M_{200}}$. The shape and color of the markers indicate the method used to estimate the mass. \added{The mass for \cite{Watkins2010} is the mass within \w{300}{kpc}.} The mass for \cite{Zaritsky2019} is a 90\% lower limit.
  Uncertainty measurements for compiled data points only include statistical uncertainties and not systematic uncertainties. Our mass estimate is in good agreement with other recent results.}
  \label{fig:comparison}
\end{figure*}

We compare our estimated ${\rm M_{200}}$ to \deleted{recent }measurements from the literature in Figure \ref{fig:comparison}. We have \deleted{only}\added{primarily} included studies from 2018 until 2021, and refer the reader to \cite{Wang2020} for an excellent overview and comparison of more studies. Note that the values of ${\rm M_{200}}$ reported by these studies are also extrapolations. Thus, it may be possible that studies agree in their mass estimates at radii where data are available (e.g., \w{100}{kpc}) but disagree in the extrapolated ${\rm M_{200}}$, or disagree in their mass estimates at \w{100}{kpc} but fortuitously agree in ${\rm M_{200}}$. The studies are ordered by year of publication, with the most recent studies towards the bottom of the figure. The marker shape and color indicate the method used to estimate the mass of the Galaxy; see \cite{Wang2020} for explanations of the different methods. Recent estimates for ${\rm M_{200}}$ seem to have converged to a total mass somewhere in the range of $1.0-1.5\times 10^{12}~{\rm M_\odot}$. The vertical line and three shading bands show the median and the 68\%, 95\%, and 99.7\% credible interval for our estimates of ${\rm M_{200}}$.

Several studies, marked with $\dagger$ in Figure \ref{fig:comparison}, report virial masses with overdensities other than $\rm \Delta_{vir} = \delta_{th}\,\Omega_m = 200$ (following the notation of \citealt{Kafle2014a}), so a conversion to our definition of ${\rm M_{200}}$ is necessary. All of these studies use the definition of $\Delta_{\rm vir}$ from \cite{Bryan1998} or apply a correction following \cite{Bland-Hawthorn2016} (which itself uses definitions from \citealt{Bryan1998}). The overdensities of these studies vary slightly, ranging from $\Delta_{\rm vir} = 97$ to $\Delta_{\rm vir} = 102.5$. We assume overdensities of $\Delta_{\rm vir} = 102$ ($\delta_{\rm th} = 340$ and $\rm \Omega_m = 0.3$) for simplicity---with differences being a few percent at most---and apply a correction factor of $0.84$ to all reported ${\rm M_{vir}}$ and the associated uncertainties following \cite{Bland-Hawthorn2016}. This yields ${\rm M_{200}}$ values 16\% lower than the reported virial masses, which we can then use for direct comparison with other studies.

Note that \added{\cite{Zaritsky1989}, }\cite{Patel2018}, \cite{Watkins2019} and \cite{Deason2021} all report two masses, and thus each have two points in Figure \ref{fig:comparison}. \added{Using a statistical analysis based on the method of \cite{Little1987}, \cite{Zaritsky1989} report a mass of $\rm M_{200} = 0.93_{-0.12}^{0.41} \times 10^{12} \; M_\odot$ when satellite orbits are assumed to be radial, and $\rm M_{200} = 1.25_{-0.32}^{0.84} \times 10^{12} \; M_\odot$ when they are assumed to be isotropic. }\cite{Patel2018} report a mass of ${\rm M_{200} = 0.71_{-0.22}^{+0.19}\times 10^{12} \; M_\odot}$ (converted) when the Sagittarius Dwarf Spheroidal Galaxy is excluded from their analysis, and ${\rm M_{200} = 0.81_{-0.24}^{+0.24}\times 10^{12} \; M_\odot}$ when it is included. \cite{Watkins2019} also use two datasets in their analysis; they find ${\rm M_{200} = 1.08_{-0.40}^{+0.82} \times 10^{12} \; M_\odot}$ (converted) when only using \gaia\ kinematic data, and ${\rm M_{200} = 1.29_{-0.37}^{+0.63} \times 10^{12} \; M_\odot}$ when proper motions from the \textit{Hubble Space Telescope} (\textit{HST}) are also included. \cite{Deason2021} quote two masses, depending on whether the mass of a rigid Large Magellenic Cloud (LMC) with $\rm M = 0.15\times 10^{12}\;M_\odot$ is included; a more detailed comparison with \cite{Deason2021} will follow in this subsection given the similarities of their method to ours.

Two studies that we are particularly interested in comparing our results to are \cite{Zaritsky2019} and \cite{Deason2021}. The former applies a different analysis technique to an earlier version of the H3 dataset, and the latter applies the same distribution function used in this paper to a different set of halo stars out to a comparable distance.

\cite{Zaritsky2019} applied the timing argument to a sample of 32 halo stars with $R > 60~{\rm kpc}$ from the H3 Survey, and found with 90\% confidence that ${\rm 0.91\times 10^{12} \; M_\odot < M_{200} < 2.56\times 10^{12} \; M_\odot}$ (taking the conservative upper and lower limits). Our mass estimate is consistent with---but toward the lower end of---these limits. Within the 90\% credible interval, we find that ${\rm 0.91\times 10^{12} \; M_\odot < M_{200} < 1.28\times 10^{12} \; M_\odot}$. Promisingly, this suggests that the data are able to provide similar constraints on the mass of the Galaxy even under different analysis techniques, each of which has their own modeling assumptions.

\cite{Deason2021} applied the same distribution function as we use in this paper to a sample of 485 halo stars (excluding stars from the Sagittarius stream) from various surveys. These halo stars range in Galactocentric radius from $50-100~{\rm kpc}$, with distance errors of $5-10\%$. There are several noteworthy differences between their analysis and the one in this paper.

First, in \cite{Deason2021}, the slope of the tracer halo density is fixed at $\alpha = 4$, whereas we estimate it as a (nuisance) parameter.
Second, \cite{Deason2021} assume that there is no correlation between $v_b$ and $v_l$ (essentially the proper motions). We incorporate the correlations between positions and proper motions, which should provide stronger constraints on the mass to the extent that the true covariances are non-zero.
Third, \cite{Deason2021} reduce the 3D distribution of the velocity into a line-of-sight velocity distribution by marginalizing over $v_b$ and $v_l$. We maintain all three velocity components in our analysis.
Finally, uncertainties are only included for distances and line-of-sight velocities in the \cite{Deason2021} analysis. We include uncertainties for all six phase space parameters, but this is a minor difference given that the uncertainties in the positions are very small. In their work, the uncertainty estimation is done with a Monte Carlo procedure where distances and line-of-sight velocities are scattered 100 times, and the likelihood is calculated each time. In comparison, the design of our model allows us to directly incorporate measurement uncertainties, which are automatically propagated through to the posterior distribution. We acknowledge, however, that we make the assumption that all the measurement errors are Gaussian, which may not actually be the case.

\cite{Deason2021} also find that $\gamma$ and $\po$ are negatively correlated in their joint posterior distribution, whereas we find that they are strongly positively correlated. See Appendix \ref{appendix:correlation} for more details.

Our mass estimates are in good agreement with the masses reported in \cite{Deason2021}, both within the range of the both datasets and in the extrapolated masses. Within \w{100}{kpc}, we find a mass of ${\rm M(<100 \; kpc) = 0.69_{-0.04}^{+0.05} \times 10^{12} \; M_\odot}$, which is roughly 10\% higher than \cite{Deason2021}'s estimate of $\rm {M(<100 \; kpc)} = 0.63 \, \pm \, 0.03 (stat.) \, \pm \,0.13 (sys.) \times 10^{12} \; M_\odot$. Our ${\rm M_{200} = 1.08_{-0.11}^{+0.12}\times 10^{12}\;M_\odot}$ estimate lies in between \cite{Deason2021}'s post-LMC infall estimate of ${\rm M_{200} = 1.20\,\pm\,0.25\times 10^{12}\;M_\odot}$ and pre-LMC infall estimate of ${\rm M_{200} = 1.05\,\pm\,0.25\times 10^{12}\;M_\odot}$. Interestingly, \cite{Deason2021} find that including Sgr stars in their analysis biases mass estimates low; we find the opposite, with our analysis \textit{excluding} Sgr stars having a significantly lower mass estimate. However, our definition of Sgr is not limited to the Sgr stream, making it different from the definition used by \cite{Deason2021}. See Section \ref{sec:discussion-sgr} for more details.

% \needspace{4\baselineskip}
\subsection{Posterior Predictive Checks}\label{sec:discussion-ppc}

Bayesian models are flexible in that we can use data to constrain parameters as we have already done, but we can also use parameters to simulate data. We can draw samples from the joint posterior distribution of $\alpha$, $\beta$, $\gamma$, and $\po$, which were estimated with our observed data, and push those samples through the distribution function to obtain simulated Galactocentric phase-space information.

Statistically, the distribution we draw from is known as the posterior predictive distribution, which is written as
\begin{align}\label{eqn:posterior-predictive-distribution}
  p(\tilde{y}\, |\, y) = \int p(\tilde{y}\, |\, \boldsymbol{\theta})\, p(\boldsymbol{\theta}\, |\, y)\, d\boldsymbol{\theta},
\end{align}
where $\tilde{y}$ is a simulated draw (vector) from $p(\tilde{y}\, |\, \boldsymbol{\theta})$, $\boldsymbol{\theta}$ is a draw of parameters from the posterior distribution $p(\boldsymbol{\theta}\, |\, y)$, and $y$ are the observed data \citep{Gelman2013}. These simulated data $\tilde{y}$ can be plotted on top of the real data to see whether the model reasonably describes the data; this is a primarily qualitative check.

For our particular model, we draw from the posterior predictive distribution as follows. We make one draw of $\boldsymbol{\theta} = (\alpha, \beta, \gamma, \po)$ from the joint posterior distribution. The value of $\alpha$, together with a minimum radius of $r_{min} = 50~{\rm kpc}$, is used to parameterize a Pareto distribution, from which a Galactocentric distance $r$ is drawn. Then, we use the simulated $r$ and the same $\boldsymbol{\theta}$ to draw radial and tangential velocities $v_r$ and $v_t$ from the posterior predictive distribution, which is the phase-space distribution function multiplied by $v_t$.

\begin{figure}[tbh!]
  \centering
  \includegraphics[width=1\linewidth]{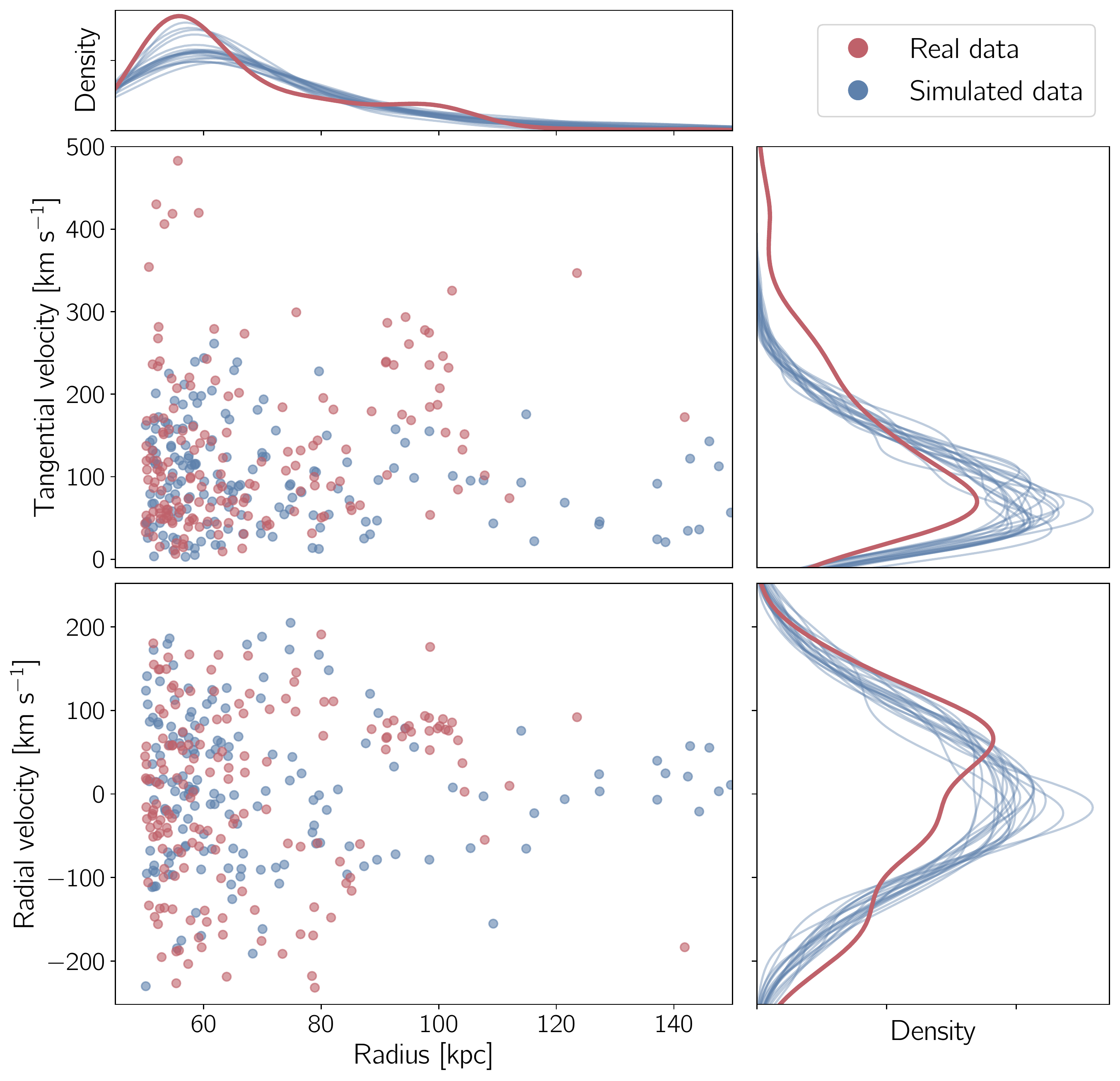}
  \caption{A plot of simulated distances, radial velocities, and tangential velocities compared to the actual data. In all panels, red corresponds to the real H3 data and blue corresponds to simulated draw drawn from the posterior predictive distribution. The two large panels are scatterplots of $r-v_r$ and joint $r-v_t$ data. The top panel shows the marginal distributions of the real distances, along with the distributions of 25 sets of simulated distances. The two panels on the right are similar to the panel on the top, and show the real and simulated distributions of the radial and tangential velocities. The model does not capture the (possibly fake) heavy tail in the distribution of tangential velocities, and also produces a heavy tail in distribution of Galactocentric radii that is not reflected in the real data.}
  \label{fig:ppc}
\end{figure}

In Figure \ref{fig:ppc}, the two large panels show the joint $r-v_r$ and joint $r-v_t$ distributions for both real and simulated data. The number of points simulated here is equal to the number of data points (168). The panels on the top and the right of the plot show the marginal posterior predictive distributions for $r$, $v_r$, and $v_t$. In each of these panels, the red line corresponds to the distribution of the real data, and the blue lines correspond to the distribution of the simulated data. Each panel has 25 blue lines, which correspond to the densities of 25 random simulations. For each simulation, 168 points are drawn from the posterior predictive distribution, as before, and the density of those points is plotted as a single line.

This plot is a valuable diagnostic that can indicate to us where the model falls short. Very roughly, the simulated data appear to match the real data reasonably well. However, upon closer inspection, there are several problems that become apparent. The first is that the distributions of simulated $r$'s do not match the distribution of $r$ for the real data very well. The second is that the heavy tail in the distribution of the real $v_t$'s is not captured by the simulated distributions. Finally, there is the asymmetry in the distribution of $v_r$, which we will further discuss in Section \ref{sec:discussion-ppc-vr}.

\needspace{4\baselineskip}
\subsubsection{The distribution of $r$}\label{sec:discussion-ppc-r}

We first examine the distribution of the radius $r$. Note that Figure \ref{fig:ppc} is plotted with a maximum radius of \w{150}{kpc}, which is roughly the maximum radius of the stars in our sample. However, some simulated points have larger values of $r$ (beyond what is shown here) because the distribution function models the slope of the stellar halo as a Pareto distribution, which has non-zero probability even at large $r$. This is not entirely realistic because it does not account for the magnitude limit of surveys. The H3 Survey in particular has a limit of $r = 18$, and we have no stars beyond approximately $140~{\rm kpc}$, but this drop-off is not reflected in the distribution function.

\begin{figure*}[htb!]
  \centering
  \includegraphics[width=1\linewidth]{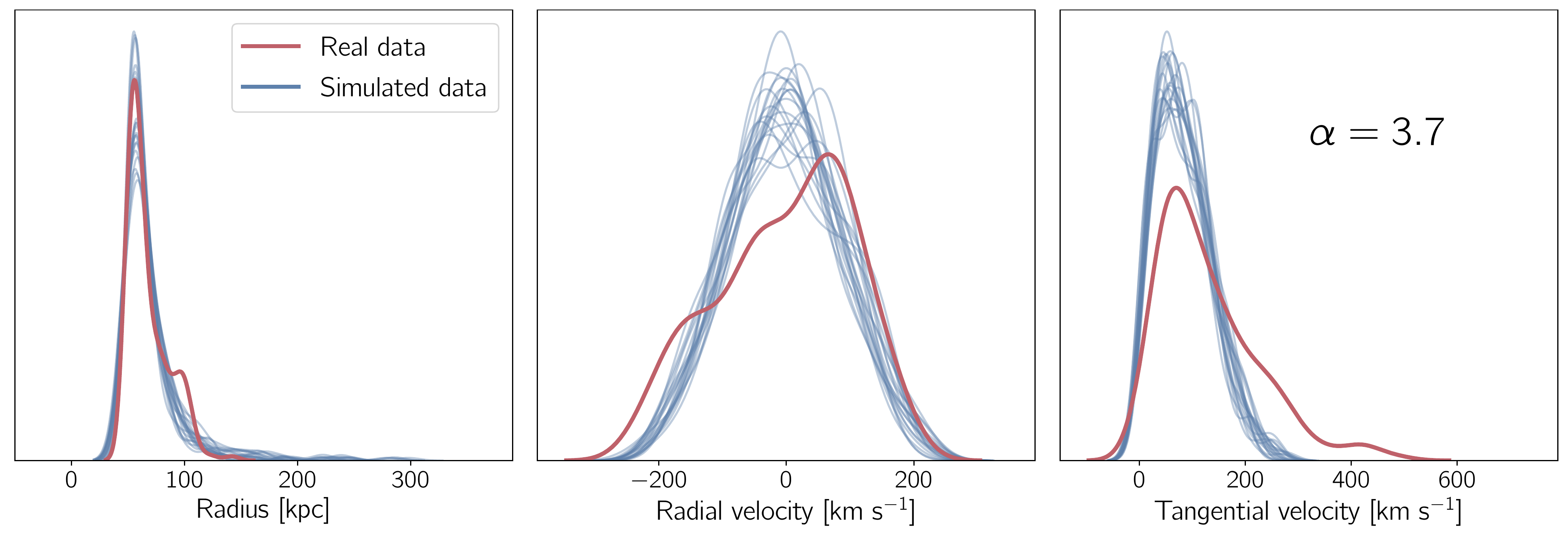}
  \caption{A plot of simulated distances, radial velocities, and tangential velocities compared to the actual data. In all panels, red corresponds to the real H3 data and blue corresponds to simulated draw drawn from the posterior predictive distribution. Each panel has 25 blue lines, corresponding to the distributions of 25 simulated sets of data points. Here, a value of $\alpha = 3.7$---specifically chosen to match the distribution of Galactocentric radii---together with random draws of $\beta$, $\gamma$, and $\po$ from the posterior distribution, is used to simulate data.}
  \label{fig:ppc-a37}
\end{figure*}

This problem is potentially exacerbated by the low estimate of $\alpha$, which results in a shallower Pareto distribution. Indeed, the simulated data do not have as sharp a ``peak'' in the density as is seen in the distribution of the real data. We can confirm that this is a result of the low $\alpha$ estimate by replacing the values for $\alpha$ in the posterior predictive simulation. Rather than drawing $\alpha$ from the posterior distribution, we can simply put in whatever value of $\alpha$ we would like to use, while still drawing $\beta$, $\gamma$, and $\po$ from the posterior distribution. We find that for a higher value of $\alpha = 3.7$ (which is roughly the maximum likelihood estimate when fitting a Pareto distribution directly to the $r$'s of the real data), the simulated distributions match the real distribution far better. This is shown in the left panel of Figure \ref{fig:ppc-a37}. However, this comes at the cost of the simulated $v_t$'s being a very poor match to the real $v_t$'s.

% \needspace{4\baselineskip}
\subsubsection{The distribution of $v_t$}\label{sec:discussion-ppc-vt}

From Figure \ref{fig:ppc}, we can see that the marginal distribution of the real $v_t$'s has a heavy tail that goes out to $\sim 500~{\rm km\,s^{-1}}$. This is not reflected in the simulated data, for which the density drops off at $\sim 300~{\rm km\,s^{-1}}$. There is a possiblity that the stars with large $v_t$'s are not real; if they were real we would expect a similar number of stars with high $v_r$'s, but we do not. The large $v_t$'s could be a result of systematically biased distances.

As the model tries its best to describe this heavy tailed data, our mass estimates may be skewed. This is due to limitations in the distribution function that we are using. We address this concern in Section \ref{sec:discussion-sextans}. On the other hand, the failure to capture this heavy tail may also be problematic. 

\begin{figure*}[htb!]
  \centering
  \includegraphics[width=1\linewidth]{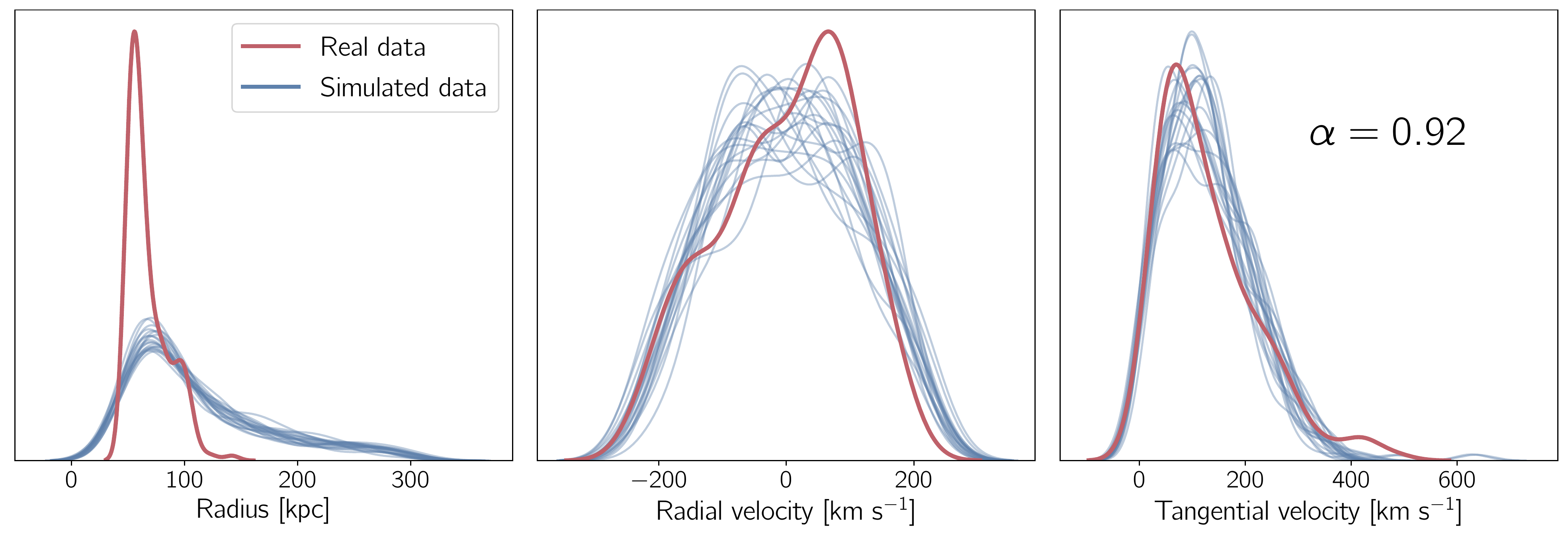}
  \caption{A plot of simulated distances, radial velocities, and tangential velocities compared to the actual data. In all panels, red corresponds to the real H3 data and blue corresponds to simulated draw drawn from the posterior predictive distribution. Each panel has 25 blue lines, corresponding to the distributions of 25 simulated sets of data points. Here, a value of $\alpha = 0.92$---specifically chosen to best match the distribution of tangential velocities---together with random draws of $\beta$, $\gamma$, and $\po$ from the posterior distribution, is used to simulate data.}
  \label{fig:ppc-a092}
\end{figure*}

In a reversal from the previous section, the real distribution of $v_t$ would be better described if we simulated data using a \textit{lower} value of $\alpha$. Indeed, the right panel of Figure \ref{fig:ppc-a092} shows that with $\alpha = 0.92$---and still using random draws of $\beta$, $\gamma$, and $\po$ from the posterior distribuion as before---simulated $v_t$ distributions match the real $v_t$ distribution very well. There are two clear problems with this. The first is that the distribution function places a functional lower bound of 3 on $\alpha$. Thus, as the $v_t$ data favor a value below 3, the posterior for $\alpha$ becomes heavily skewed and piles up against the boundary, as we observe in Figure \ref{fig:h3_pairs}. The second problem is clear from the discussion in Section \ref{sec:discussion-ppc-r} and from the left panel of Figure \ref{fig:ppc-a092}; such a low value of $\alpha$ results in simulated $r$ that match the real $r$ very poorly. 

In summary, the distribution of $v_t$ favors a lower value of $\alpha$, while the distribution of $r$ favors a higher value of $\alpha$. Thus, $\alpha$ is both pulled towards higher values to better match the distribution of $r$ and towards lower values to better match the distribution of $v_t$. This inability to simulataneously fit the distributions of $r$ and $v_t$ reflects a limitation of the distribution function.

\needspace{4\baselineskip}
\subsubsection{The distribution of $v_r$}\label{sec:discussion-ppc-vr}

In Figure \ref{fig:ppc} (and in Figures \ref{fig:ppc-a37} and \ref{fig:ppc-a092}), the simulated distributions of $v_r$ generally appear to be a fairly good match to the real distribution of $v_r$. However, there is asymmetry in the real distribution---caused by the large clump of stars at $v_r \simeq 100~{\rm km\,s^{-1}}$---that is not captured by most of the simulated distributions. However, in the lower right panel of Figure \ref{fig:ppc}, there does appear to be one simulated distribution that does have similar asymmetry in the opposite direction (i.e., the ``peak'' is toward negative radial velocities). We thus conclude that it is unlikely---but possible---that the observed asymmetry is consistent with random fluctuations.

This clump of stars contains numerous tracers that are all at a similar distance from the Galactic center ($\sim 100~{\rm kpc}$) and they also have similar radial velocities ($\sim 90~{\rm km\,s^{-1}}$). This gives us reason to believe that they are not independent in phase space, and rather are part of some substructure in the halo. We dedicate Section \ref{sec:discussion-substructure} to analyzing the effect of substructure on our mass estimates.

% \needspace{5\baselineskip}
\subsection{Impact of Substructure on Mass Estimates}\label{sec:discussion-substructure}

The distribution function method is an equilibrium-based method that assumes that tracers are independent. However, it is clear from Figure \ref{fig:ppc} that this is not the case;
in particular, note the clump of Sextans stars at $r\simeq 100~{\rm kpc}$, which have similar radial velocities. These Sextans stars are more clearly highlighted in Figure \ref{fig:lze}. Other large-scale structures like the Sagittarius stream \citep{Ibata2001, Belokurov2006} might also be of concern, given that recent studies have found that the presence of substructure can lead to biased mass estimates \citep[e.g.,][]{Grand2019, Erkal2020, Deason2021}. \cite{Cunningham2019} have also found that the velocity anisotropy can vary with the position in the sky, which if true, may present problems for our analysis that assumes a single anisotropy that varies neither with position in the sky nor with radius from the Galactic center. To investigate the effect of substructure on our mass estimates, we select various subsets of data and rerun our analysis on each subset. We clarify here that by ``substructure'', we are referring to cold, correlated structure that is dynamically unmixed.

% \needspace{5\baselineskip}
\subsubsection{Split-Sky Tests}\label{sec:discussion-splitsky}

We explore whether spatially coherent substructure in the sky can affect our estimates of the mass (and the velocity anisotropy). We perform split-sky tests, whereby a subset of stars are removed from the full set based on their positions in the sky. We make four cuts based on right ascension. In each cut, a quarter of the sky is removed, with the expectation that large-scale structures are broken up or entirely excluded in certain subsets.
Table \ref{table:splitsky} shows the results of applying this splitting to the H3 dataset.

\begin{deluxetable*}{cccccccc}[!ht]
  \tablecaption{Results from split-sky tests. \label{table:splitsky}}
  \tablehead{
    \colhead{Cut} & \colhead{Stars} & \colhead{$\alpha$} & \colhead{$\beta$} & \colhead{$\gamma$} & \colhead{$\po$} & \colhead{$r_{200}$} & \colhead{${\rm M_{200}}$} \\
    \colhead{} & \colhead{Number} & \colhead{} & \colhead{} & \colhead{} & \colhead{$10^{4}~{\rm km^{2}\,s^{-2}}$} & \colhead{$\rm kpc$} & \colhead{$\times 10^{12}~{\rm M_\odot}$}
  }
  \startdata
  $0^\circ - 90^\circ$ removed    & 148 & $3.00_{-0.00}^{+0.00}$ & $0.40_{-0.06}^{+0.05}$ & $0.43_{-0.05}^{+0.05}$ & $51.09_{-5.86}^{+6.51}$ & $216.37_{-7.63}^{+7.84}$ & $1.08_{-0.11}^{+0.12}$ \\
  $90^\circ - 180^\circ$ removed  & 122 & $3.00_{-0.00}^{+0.00}$ & $0.37_{-0.04}^{+0.04}$ & $0.47_{-0.06}^{+0.05}$ & $49.93_{-6.89}^{+6.99}$ & $205.07_{-7.31}^{+7.62}$ & $0.92_{-0.09}^{+0.11}$ \\
  $180^\circ - 270^\circ$ removed & 107 & $3.00_{-0.00}^{+0.00}$ & $0.42_{-0.06}^{+0.06}$ & $0.44_{-0.05}^{+0.05}$ & $60.13_{-6.63}^{+6.81}$ & $230.64_{-8.40}^{+8.97}$ & $1.31_{-0.14}^{+0.16}$ \\
  $270^\circ - 360^\circ$ removed & 127 & $3.00_{-0.00}^{+0.00}$ & $0.34_{-0.06}^{+0.06}$ & $0.43_{-0.06}^{+0.06}$ & $51.37_{-7.02}^{+8.05}$ & $217.18_{-8.06}^{+8.21}$ & $1.09_{-0.12}^{+0.13}$ \\
  \enddata
  \tablecomments{Estimated values are the medians of the posterior distributions. The uncertainties give the 16th and 84th percentiles.}
\end{deluxetable*}

We find that all the split-sky tests apart from the one where the stars with right ascension of $180^\circ - 270^\circ$ were removed, the estimated parameters and mass are in good agreement with each other and with the results from the full set. For the run where stars with right ascension of $180^\circ - 270^\circ$ were removed, the estimated mass is higher than the mass from other runs due to the gravitational potential having a larger scale factor and a shallower slope. This affects not only the mass estimate at a given radius, but also the estimated virial radius. The largest difference in the virial radii and the masses between the runs looks to be roughly $30\%$, which is larger than the statistical uncertainties in the estimates. This suggests that to obtain more reliable estimates of the true mass, it is not the precision of our estimates that we should focus on, but rather systematic effects.

% \needspace{5\baselineskip}
\subsubsection{Sextans stars}\label{sec:discussion-sextans}

\begin{figure}[htb!]
  \centering
  \includegraphics[width=\linewidth]{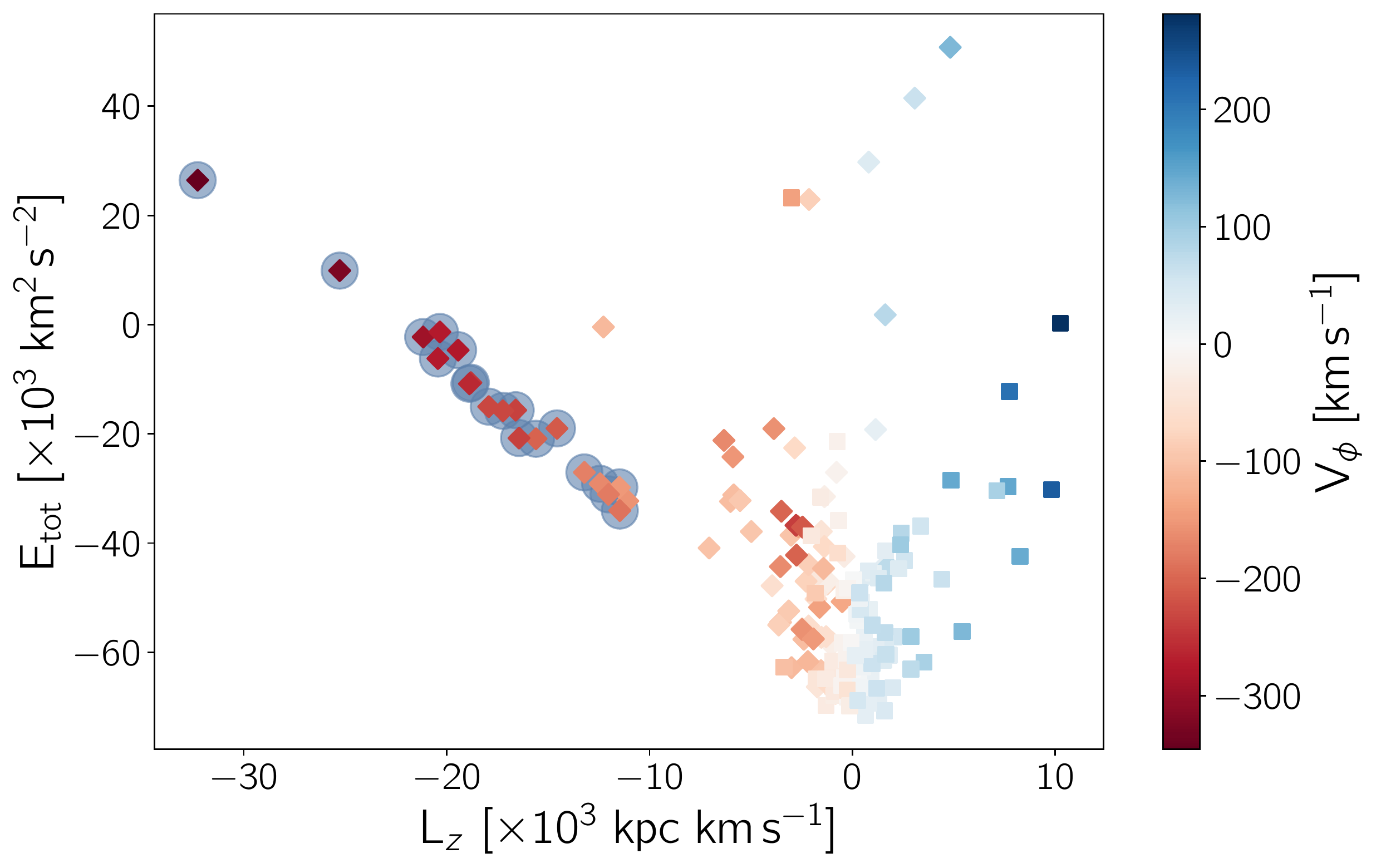}
  \caption{Full set plotted in $L_z-E$ space, colored by $V_\phi$. Stars associated with Sgr are plotted as diamonds, and stars not associated with Sgr are plotted as squares. Sextans stars, toward the left of the plot, are circled in blue. They have much larger angular momenta than most of the other stars.}
  \label{fig:lze}
\end{figure}

Figure \ref{fig:lze} shows the $L_z-E$ plot of the full set. The most apparent outliers in the figure are the stars associated with the Sextans dwarf galaxy, circled in blue. These stars have a right ascension of $\alpha \simeq 153^\circ$, and have angular momenta that are clearly separated from the bulk of the other stars.
Many of these stars also make up the clump of stars with $\sim R = 100~{\rm kpc}$ and $\sim V_r = 80~{\rm km\,s^{-1}}$ in the lower left panel of Figure \ref{fig:ppc}.
We identify 19 stars.

To investigate whether these stars have a significant impact on our mass estimates, we perform two tests. In the first test, we remove these stars from the analysis. In the second test, we average the properties of the stars; this essentially de-weights them in the analysis. The latter is done by averaging the properties (distances, velocities, etc.) of stars with angular momentum between $L_z = -10\times 10^{3}~{\rm kpc\,km\,s^{-1}}$ and $L_z = -17\times 10^{3}~{\rm kpc\,km\,s^{-1}}$ ($n=9$) into one data point, and the stars with angular momentum less than $L_z = -17\times 10^{3}~{\rm kpc\,km\,s^{-1}}$ ($n=10$) into another.

For the first run, we find a median mass estimate of $\rm M_{200} = 1.00_{-0.11}^{+0.11}\times 10^{12} \; M_\odot$ and an anisotropy of $\beta = 0.39_{-0.04}^{+0.03}$. For the second run, we find a median mass estimate of $\rm M_{200} = 1.01_{-0.10}^{+0.11}\times 10^{12} \; M_\odot$ and an anisotropy of $\beta = 0.39_{-0.04}^{+0.04}$. The results of these tests indicate that removing the Sextans stars leads to a higher estimate for $\beta$, indicating more radial orbits. This makes sense given their large (absolute) values of $L_z$. The mass estimates in both cases are lower than the estimate using the full sample, but still within \deleted{$1\,\sigma$}\added{the 68\% credible interval}. These tests show that despite being so separated from the rest of the stars in phase space, the Sextans stars do not seem to have a statistically significant impact on the estimated parameters.

% \needspace{3\baselineskip}
\subsubsection{Sagittarius Cut}\label{sec:discussion-sgr}

We also perform a run with Sagittarius stars excluded. This is done by removing stars in the full set which have a flag of \verb|Sgr_FLAG=1|. This flag, loosely, marks stars associated with Sagittarius by making a cut in $L_y-L_z$ space \citep{Johnson2020}. This flag potentially includes stars that are not actually associated with Sagittarius. The resulting sample with Sagittarius stars excluded has $n=84$ stars; the diamonds in Figure \ref{fig:lze} show the $L_z-E$ distribution of Sgr stars. See \cite{Johnson2020} for more details about Sgr in H3.

We find that excluding the Sagittarius stars results in more radial orbits---again, as expected---with $\beta = 0.44_{-0.04}^{+0.05}$.
The median mass estimate is ${\rm M_{200} = 0.76_{-0.09}^{+0.11}\times 10^{12}\; M_\odot}$, a 30\% decrease from the estimate using the full set.
This difference is larger than the $\sim 15\%$ variation seen in the split-sky tests of Section \ref{sec:discussion-splitsky}, and shows that structure may not be localized on the sky. However, because the Sgr flag makes a cut in phase space, the assumptions of the phase space distribution function are violated in this analysis. Thus, this uncertainty is likely not representative of the true (systematic) uncertainty in the analysis.

  \needspace{5\baselineskip}
  \section{Summary and Future Prospects}\label{sec:conclusion} %%%%%%%%%%%%%%%%%%%%%%%%%%%%%%%%%%

  \needspace{2\baselineskip}
\subsection{Summary}
In this study, we estimate the mass distribution of the Milky Way halo. To this end, we extend the Bayesian multilevel model of \cite{Eadie2017} and \cite{Eadie2019} to include a full probabilistic treatment of all phase-space parameters and to incorporate correlation information between these parameters. Together with the increase in the number of tracers used relative to previous studies (\citealt{Eadie2019} and \citealt{Slizewski2021}), this introduces a significant number of additional parameters that need to be estimated.

To make this computation faster, we replace the inference engine with NUTS by rewriting the codebase in the probabilistic programming language Stan, resulting in several orders of magnitude of speedup compared to GME. We use 168 halo stars with distances out to \w{140}{kpc} as tracers for the gravitational potential of the halo. Positions and proper motions for the stars are provided by the \gaia\ satellite, and distances and radial velocities are provided by the H3 Survey. We find a median estimate for the total mass of the Milky Way of ${\rm M_{200} = 1.08_{-0.11}^{+0.11}\times 10^{12} \; M_\odot}$, in good agreement with recent studies. We also find that the orbits in the outer halo are slightly radial, with anisotropy $\beta = 0.35_{-0.05}^{+0.06}$.

To test the validity of our results, we perform posterior predictive checks and sensitivity analyses with a focus on substructure. We find that the single power-law slope fails to simultaneously capture the distribution of $r$ and $v_t$ of the tracer population. Furthermore, the distribution function does not take into account the magnitude limit in the H3 Survey, leading to a heavier tail in the distribution of distances than is observed in the real data. Finally, we find that the presence of substructure introduces roughly 15\% uncertainty into our estimate, which is larger than the statistical errors in our analysis.

\needspace{5\baselineskip}
\subsection{Future prospects}

We recommend that future work considering the use of RWM for Bayesian inference look into using NUTS instead, especially in the case of high-dimensional problems.

There are many avenues that follow-up work could take. 
With NUTS as a powerful inference engine, our model will be suitable for use with data from (combinations of) large-scale spectroscopic surveys like DESI \citep{DESICollaboration2016, Prieto2020} and SDSS-V \citep{Kollmeier2017}. 

Based on simulations, we find that our estimated $\rm M_{200}$ may be a slight underestimate, but this is based on a single simulated galaxy, and more work looking into mock data where the true mass profiles are available would be valuable. Our estimate of $\rm M_{200}$ carries with it significant uncertainty due to extrapolation, which is a problem that is present in most studies estimating the mass of the Galaxy. Future surveys providing kinematic information for stars out to the virial radius would eliminate (or at least reduce) the need for extrapolations, thereby reducing uncertainty in virial mass estimates.

With respect to the method used in this work, many extensions could also be made. One area that could be improved extensively is the exact distribution function. We make a number of possibly unrealistic assumptions in our current distribution function for the sake of simplicity, including spherical symmetry of the halo, a single tracer population, and a single fixed anisotropy for all distances and positions. It may be worthwhile to devise more flexible distribution functions---possibly even ones that are numerically evaluated, given how fast inference with NUTS is---that allow for an anisotropy that varies with radius, multiple independent tracer populations, or triaxial potentials \citep{Binney1987}. \cite{Green2020} have also shown that normalizing flows can be used as incredibly flexible distribution functions that do not require analytic models and only rely on minimal physical assumptions.

The statistical model itself could be extended to include, for example, an indicator variable for whether a tracer is bound to the galaxy, which would perhaps allow it to deal with extreme velocity stars more effectively. However, equilibrium methods still assume independence between tracers; the development of methods that can better deal with substructure and non-independence of tracers (possibly by identification and removal) would be very valuable for reducing systematics in Milky Way mass estimates. Exploring methods for estimating the mass without relying on dynamical tracers would also be worthwhile (see, e.g., \citealt{Zaritsky2017} and \citealt{Craig2021}).

More generally, simulations to better characterize the Galaxy and its halo would allow us understand whether our modeling assumptions are valid. This is particularly important now---as the quality and quantity of data increase and statistical uncertainties are reduced, systematics are the dominant uncertainty that need to be dealt with.

  \section*{Acknowledgements}

  \added{We thank the anonymous referee for the careful reading of this manuscript and for providing helpful suggestions.}
  JS thanks the Dunlap Institute, which is funded through an endowment established by the David Dunlap family and the University of Toronto.
  GME acknowledges funding from NSERC through Discovery Grant RGPIN-2020-04554 and from UofT through the Connaught New Researcher Award, both of which supported this research.
  YST is grateful to be supported by the NASA Hubble Fellowship grant HST-HF2-51425.001 awarded by the Space Telescope Science Institute. 
  YST acknowledges financial support from the Australian Research Council through DECRA Fellowship DE220101520.
  We thank the Hectochelle operators Chun Ly, ShiAnne Kattner, Perry Berlind, and Mike Calkins, and the CfA and U. Arizona TACs for their continued support of the H3 Survey. 

  This work has made use of data from the European Space Agency (ESA) mission {\it Gaia} (\url{https://www.cosmos.esa.int/gaia}), processed by the {\it Gaia} Data Processing and Analysis Consortium (DPAC, \url{https://www.cosmos.esa.int/web/gaia/dpac/consortium}). Funding for the DPAC has been provided by national institutions, in particular the institutions participating in the {\it Gaia} Multilateral Agreement.

  \facilities{
    MMT (Hectochelle),
    \gaia
  }

  \software{
    Stan \citep{Hoffman2014, Carpenter2017},
    NumPy \citep{Harris2020},
    Matplotlib \citep{Hunter2007},
    Seaborn \citep{Waskom2021},
    Daft \citep{Foreman-Mackey2019}
    Astropy \citep{Astropy2013, Astropy2018}
  }

  %%%%%%%%%%%%%%%%%%%%%%%%%%%%%%%%%%%
  \newpage
  \bibliography{sources}

  %%%%%%%%%%%%%%%%%%%%%%%%%%%%%%%%%%
  \appendix

  \needspace{5\baselineskip}
\section{Position and velocity transformations}\label{appendix:transformations}

We first define several constants, vectors, and matrices that will be used for the transformations. These are taken from the Galacticentric frame in Astropy \citep{Astropy2013, Astropy2018}:
\begin{itemize}
  \item RA of the Galactic Center: $\alpha_{GC} = 266.4051^\circ$
  \item Dec of the Galactic Center: $\delta_{GC} = -28.936175^\circ$
  \item distance from the Sun to the Galactic Center: $d_{GC} = 8.122~{\rm kpc}$
  \item height of the Sun above the Galactic plane: $z_\odot = 0.0208~{\rm kpc}$
  \item roll angle (for aligning the z-axis with the Galactic y-z plane): $\eta = 58.5986320306^\circ$
  \item motion of the sun around the Galaxy in ${\rm km\,s^{-1}}$: $v_\odot = \begin{pmatrix}
      12.9 \\
      245.6 \\
      7.78
    \end{pmatrix}$
  \item a clockwise rotation matrix of $-\delta_{GC}$ about the y-axis: $R_1 = \begin{bmatrix}
      \cos{\delta_{GC}} & 0 & \sin{\delta_{GC}} \\
      0 & 1 & 0 \\
      -\sin{\delta_{GC}} & 0 & \cos{\delta_{GC}}
    \end{bmatrix}$
  \item a clockwise rotation matrix of $\alpha_{GC}$ about the z-axis: $R_3 = \begin{bmatrix}
      \cos{\delta_{GC}} & \sin{\delta_{GC}} & 0 \\
      -\sin{\delta_{GC}} & \cos{\delta_{GC}} & 0 \\
      0 & 0 & 1
    \end{bmatrix}$
  \item a clockwise rotation matrix of $\eta$ about the x-axis: $R_3 = \begin{bmatrix}
      1 & 0 & 0 \\
      0 & \cos{\delta_{GC}} & \sin{\delta_{GC}} \\
      -\sin{\delta_{GC}} & 0 & \cos{\delta_{GC}}
    \end{bmatrix}$
  \item full rotation matrix: $R = R_3 R_1 R_2$
  \item angle to account for height of the Sun above the Galactic plane: $\theta = \arcsin(z_\odot / d_{GC})$
  \item clockwise rotation matrix of $\theta$ about the y-axis: $H = \begin{bmatrix}
      \cos{\theta} & 0 & \sin{\theta} \\
      0 & 1 & 0 \\
      -\sin{\theta} & 0 & \cos{\theta}
    \end{bmatrix}$
\end{itemize}

Observed positions $\alpha$, $\delta$, $d$ are in spherical heliocentric coordinates. To make the transformation to Cartesian Galactocentric coordinates $X_{GC}$, $Y_{GC}$, $Z_{GC}$, we first convert them to Cartesian heliocentric positions as follows:
\begin{align}
  x_{ICRS} &= d \cos{\alpha} \cos{\delta} \\
  y_{ICRS} &= d \sin{\alpha} \cos{\delta} \\
  z_{ICRS} &= d \sin{\delta} \\
  r_{ICRS} &= \begin{pmatrix}
    x_{ICRS} \\
    y_{ICRS} \\
    z_{ICRS}
  \end{pmatrix}
  \end{align}

  These Cartesian heliocentric positions are then transformed to Cartesian Galactocentric ones using the matrices defined above:
  \begin{align}
    r_{GC} = H(R r_{ICRS} - d_{GC} \hat{x}_{GC}) = \begin{pmatrix}
      X_{GC} \\
      Y_{GC} \\
      Z_{GC}
    \end{pmatrix}
  \end{align}
  where $d_{GC} \hat{x}_{GC} = (d_{GC}, 0, 0)$ is an offset along the x-axis to account for the distance to the Galactic Center.

  The velocities $\mu_\alpha$, $\mu_\delta$, $v_{los}$ need to be transformed into spherical Galactocentric coordinates. To do this, we again need to first transform them to Cartesian heliocentric velocities:
  \begin{align}
    V_{X,ICRS} &= v_{los} \cos{\alpha} \cos{\delta} - \mu_\alpha \sin{\alpha} - \mu_\delta \cos{\alpha} \sin{\delta} \\
    V_{Y,ICRS} &= v_{los} \sin{\alpha} \cos{\delta} + \mu_\alpha \cos{\alpha} - \mu_\delta \sin{\alpha} \sin{\delta} \\
    V_{Z,ICRS} &= v_{los} \sin{\delta} + \mu_\delta \cos{\delta} \\
    V_{ICRS} &= \begin{pmatrix}
      V_{X,ICRS} \\
      V_{Y,ICRS} \\
      V_{Z,ICRS}
    \end{pmatrix}
    \end{align}

    We then convert these velocities to Cartesian Galactocentric coordinates using the same matrices as above, while also accounting for the motion of the Sun in the Galaxy:
    \begin{align}
      V_{XYZ,GC} = H \, R \, V_{ICRS} + v_\odot = \begin{pmatrix}
        V_{X,GC} \\
        V_{Y,GC} \\
        V_{Z,GC}
      \end{pmatrix}
    \end{align}

    Finally, we convert the Cartesian Galactocentric coordinates to spherical Galactocentric coordinates. This requires that we have Cartesian Galactocentric positions already. We first define the projected distance as
    \begin{align}
      r_{proj} = \sqrt{X_{GC}^2 + Y_{GC}^2}.
    \end{align}
    We then convert the Cartesian positions into spherical positions as follows:
    \begin{align}
      \begin{pmatrix}
        d \\
        \theta \\
        \phi \\
      \end{pmatrix}
      =
      \begin{pmatrix}
        \lVert r_{GC} \rVert \\
        \arcsin{(z_\odot \, / \, \lVert r_{GC} \rVert)} \\
        \atantwo(Y_{GC}, X_{GC})\\
      \end{pmatrix}
    \end{align}
    Then the spherical Galactocentric velocities are given by:
    \begin{align}
      V_{R,GC} &= \frac{r_{GC} \, V_{XYZ,GC}}{r_{proj}^2} \\
      V_{\theta,GC} &= -d\, \frac{(Z_{GC} (X_{GC} V_{X,GC} + Y_{GC} V_{Y,GC}) - r_{proj}^2 V_{Z,GC})}{d^2 \, r_{proj}} \\
      V_{\phi,GC} &= d \cos{(\theta)} \frac{X_{GC} V_{Y,GC} - Y_{GC} V_{X,GC}}{r_{proj}^2}
    \end{align}

    \newpage
    \,

    \needspace{5\baselineskip}
    \section{Equation for the posterior distribution}\label{appendix:posterior}

    \definecolor{nordgreen}{HTML}{A3BE8C}
    \definecolor{nordblue}{HTML}{5E81AC}
    \definecolor{nordred}{HTML}{BF616A}
    \definecolor{nordpurple}{HTML}{B48EAD}

    \added{
      The full expression for the posterior distribution is:
    \begin{align}
    \textcolor{nordpurple}{p(\alpha) \, p(\beta) \, p(\gamma, \po)} \prod_i^N \textcolor{nordblue}{p( d_{i}^{obs} | d_i, \sigma_{d_i} ) \, p( v_{los,i}^{obs} | v_{los,i}, \sigma_{v_{los,i}})} \nonumber \\
    \textcolor{nordblue}{p(\alpha_i^{obs}, \delta_i^{obs}, \mu_{\alpha_i}^{obs}, \mu_{\delta_i}^{obs}, | \alpha_i, \delta_i, \mu_{\alpha_i}, \mu_{\delta_i}, \Sigma_i)} \, \textcolor{nordred}{p( T(d_i, v_{los,i}, \alpha_i, \delta_i, \mu_{\alpha_i}, \mu_{\delta_i}) | \alpha, \beta, \gamma, \po )}, 
    \end{align}
    where the first three terms (in purple) are the priors, the next three terms (in blue) make up the measurement error model, and the last term (in red) is the distribution function. Note the distinction between $\alpha$ and $\alpha_i$; the former is the (global) parameter for the distribution function, and the latter indicates the right ascension of a particular star $i$. The $\Sigma_i$ variable is the $4\times4$ covariance matrix for the position and proper motion of a star $i$. The $T$ in the distribution function term is the Heliocentric to Galactocentric position and velocity transformation (see Appendix \ref{appendix:transformations}).
  }

    \newpage
    \,

    \needspace{5\baselineskip}
    \section{Comparison of Stan results to previous results}\label{appendix:stancheck}

  As a demonstration of the capabilities of NUTS, we apply it to two datasets for which we have already obtained results using the GME code. The first is a dataset of 32 Milky Way globular clusters from \cite{Vasiliev2019} with complete 6D phase-space information, analyzed in \cite{Eadie2019}. The second dataset consists of 32 dwarf galaxies with complete data \citep[see][and references therein]{Fritz2018,Riley2019,GaiaCollaboration2020}, analyzed in \cite{Slizewski2021}.

To properly compare our results from Stan to the old results using GME, we use the same data and prior distributions that were used in these previous studies. The hyperpriors used for analyzing the globular cluster data set are given in Table \ref{table:hyperpriors-gc}\added{, and those used for the dwarf galaxy analysis are given in Table \ref{table:hyperpriors-dg}}. 

\deleted{The slope of the tracer population, $\alpha$, is parameterized by a shifted gamma distribution, which}\added{The shifted gamma distribution} has the following probability distribution function (pdf):
\begin{align}\label{eq:shifted-gamma}
  f(x \,|\, a, b, s) = \frac{b^a}{\Gamma(a)}\,(x - s)^{a - 1}\,e^{-b(x - s)},
\end{align}
where $\Gamma$ is the gamma function.

\begin{deluxetable}{ccc}[!ht]
  \tabletypesize{\footnotesize}
  \tablecaption{Hyperprior distributions used for the globular cluster analysis. \label{table:hyperpriors-gc}}
  \tablehead{
    \colhead{Model Parameter} & \colhead{Distribution} & \colhead{Distribution parameters}
  }
  \startdata
  $\alpha$ & Shifted Gamma & $a=2.993$, $b = 2.824$, $s=3$ \\
  $\beta$  & Uniform       & $a=-0.5$, $b=1$ \\
  $\gamma$ & Normal        & $\mu=0.5$, $\sigma=0.06$ \\
  $\po$    & Uniform       & $a=1$, $b=200$ \\
  \enddata
  \tablecomments{The shifted gamma distribution is parameterized with shape $a$, rate (i.e., inverse scale) $b = 1/\theta$, and shift $s$. See Equation \ref{eq:shifted-gamma} for more details.}
\end{deluxetable}

\begin{deluxetable}{ccc}[!ht]
  \tablecaption{Hyperprior distributions used for the dwarf galaxy analysis. \label{table:hyperpriors-dg}}
  \tablehead{
    \colhead{Model Parameter} & \colhead{Distribution} & \colhead{Distribution parameters}
  }
  \startdata
  $\alpha$ & Shifted Gamma & $a=67.1$, $b=133.8184$, $s=3$ \\
  $\beta$  & Uniform       & $a=-3$, $b=1$ \\
  $\gamma$ & Normal        & $\mu=0.50$, $\sigma=0.06$ \\
  $\po$    & Gamma         & $a=32.4$, $b=0.69$ \\
  \enddata
  \tablecomments{The gamma distribution is parameterized with shape $a$ and rate (i.e., inverse scale) $b = 1/\theta$. See Equation \ref{eq:shifted-gamma} for more details about the shifted gamma distribution.}
\end{deluxetable}

\begin{figure}[htb!]
  \centering
  \gridline{%
    \fig{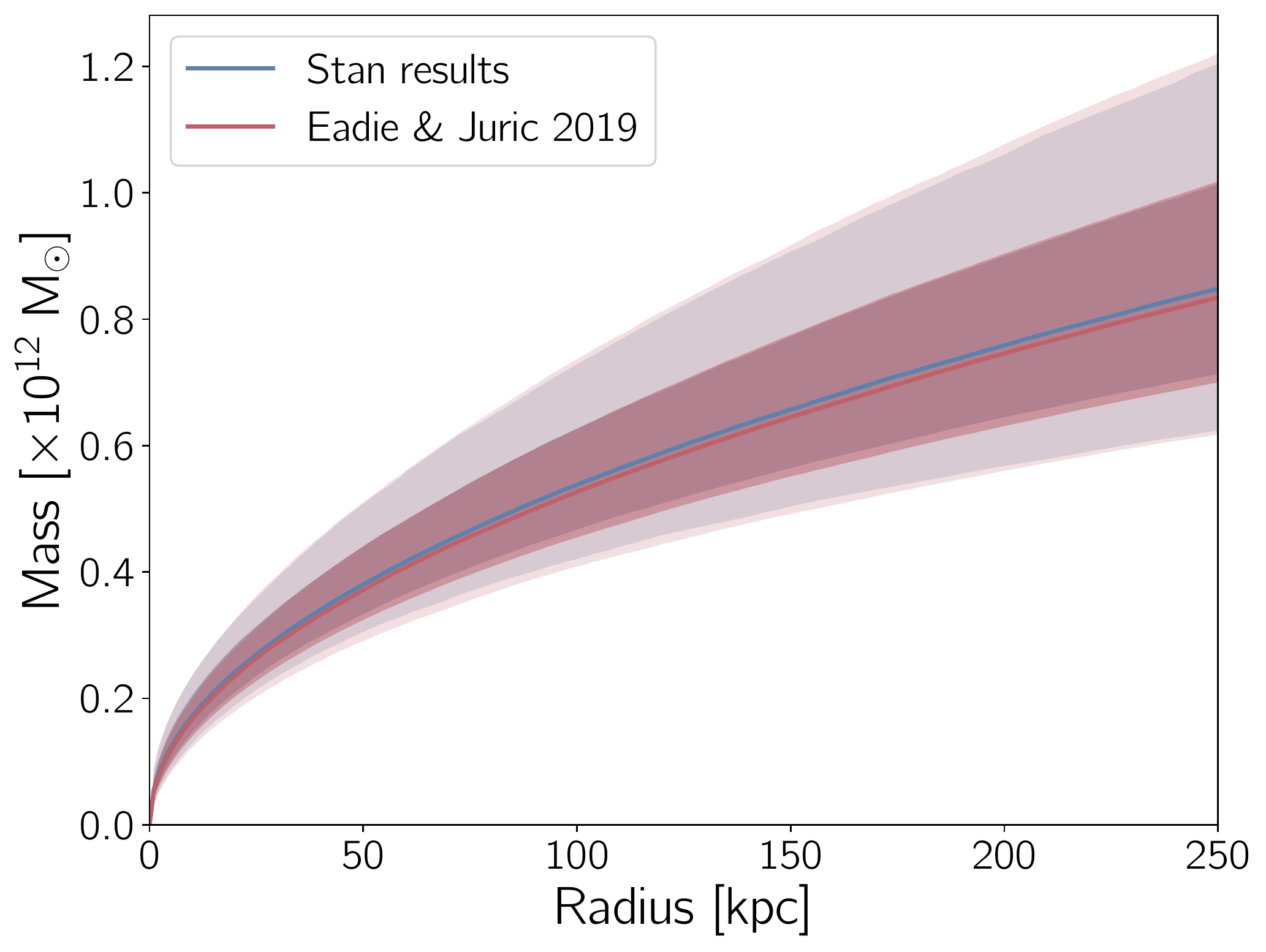}{0.45\textwidth}{(a)}
    \fig{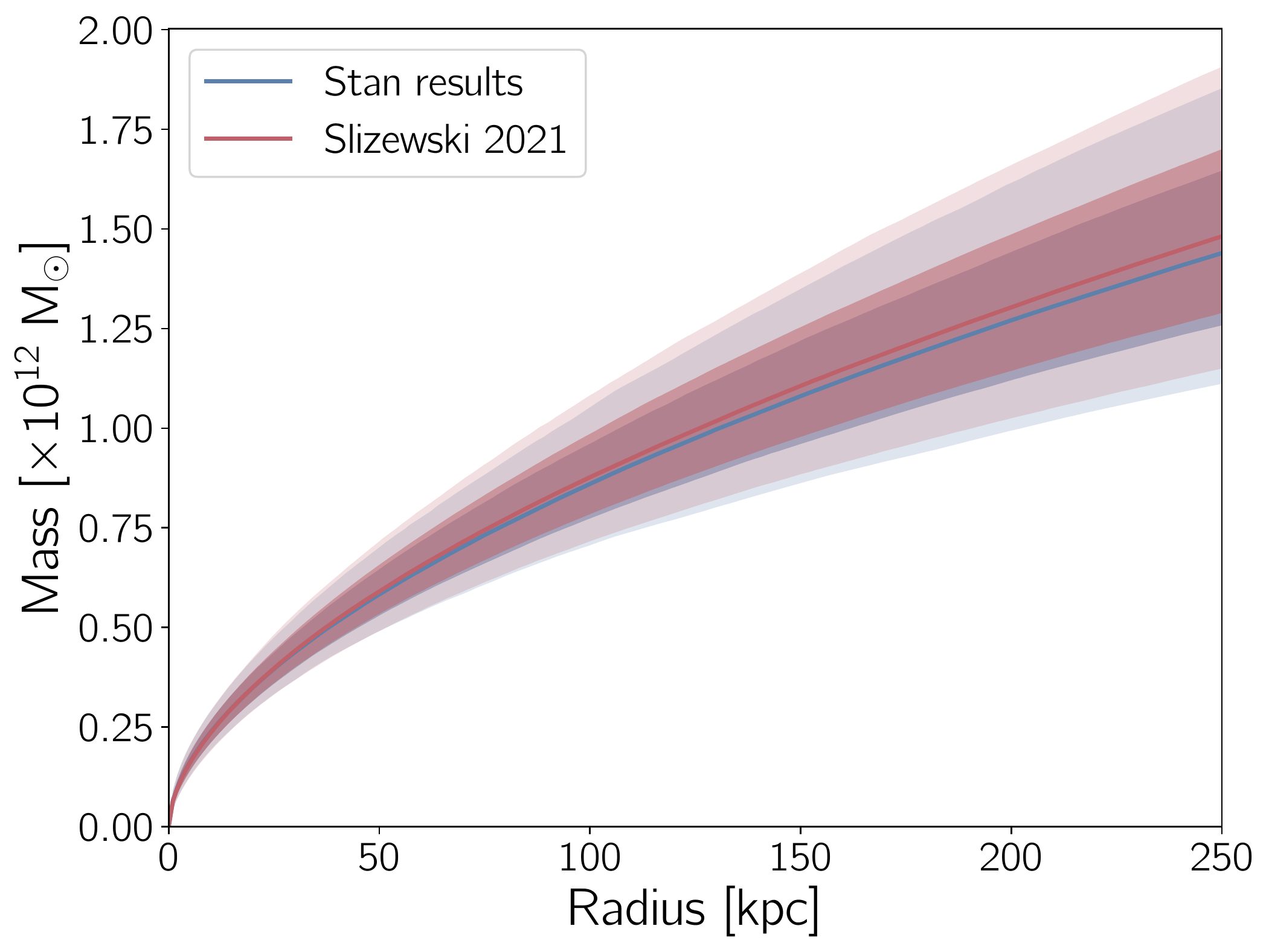}{0.45\textwidth}{(b)}
  }
  \caption{Cumulative mass distribution of the Milky Way out to \w{250}{kpc} as inferred from the kinematics of \textbf{(a)} 32 globular clusters from \cite{Vasiliev2019}, and \textbf{(b)} 32 dwarf galaxies from \cite{Fritz2018} and \cite{Riley2019}. In both panels, the lines show the median values, the dark shaded regions show the \deleted{50\%}\added{68\%} credible intervals, and the light shaded regions show the 95\% credible intervals. Included are the results of the \cite{Eadie2019} and \cite{Slizewski2021} analyses (red) and the same results as reproduced in Stan (blue); for both datasets they are in good agreement at all radii, and the Stan code was significantly faster to run.}
  \label{fig:comparison-gc-dg}
\end{figure}

The estimated cumulative mass profile\added{s for both analyses are} \deleted{is }shown in Figure \ref{fig:comparison-gc-dg}, along with the cumulative mass profile\added{s} from \cite{Eadie2019}\added{ and \cite{Slizewski2021}}. The median values are shown as the solid lines, and the \deleted{50\%}\added{68\%} and 95\% credible intervals from both analyses are shown with dark and light shading. The two codes yield masses that are in very good agreement at all radii and at different credible intervals; there is significant overlap between the red and blue shaded regions.

Although we are able to obtain the results that are consistent with \cite{Eadie2019}, the computational cost of running the new model in Stan is significantly lower than with GME (also see \citep{Eadie2015}). GME requires hours of semi-automated tuning and sampling, and the MCMC chains need to be thinned due to high autocorrelation (final three chains have a total length of $90,000$ and achieve a effective sample sizes of $\sim 2000$). In comparison, NUTS takes less than two minutes to both compile and sample. In fact, the compilation of the model takes up the bulk of this time, and the sampling only takes $\sim 30$ seconds to complete. We run four chains, each with length $2000$, with only the latter half of each chain being used for calculations (the first half of each chain are used for warmup); no thinning is necessary, and we achieve effective sample sizes of $2700-5200$ for the four main parameters.

\deleted{\cite{Slizewski2021} use the same model as described in Section \ref{sec:model-design} together with 32 dwarf galaxies from \cite{Riley2019} (see also references therein) to estimate the mass distribution of the Galaxy. The hyperpriors used for analyzing the dwarf galaxy data set are given in Table \ref{table:hyperpriors-dg}. We again reproduce this analysis with Stan both as a check to ensure that we are able to recover the same cumulative mass profile and as a demonstration of the speed of NUTS.}

\deleted{We find posteriors distributions for $\Phi_0$, $\gamma$, $\alpha$, and $\beta$ that are similar to those found by \cite{Slizewski2021}. In Figure \ref{fig:comparison-dg} we show the median and the \deleted{50\%}\added{68\%} and 95\% credible intervals for the cumulative mass profile as inferred using the new Stan code and for the analysis from \cite{Slizewski2021}. Again, the two profiles are in very good agreement, and NUTS is significantly faster.}

    \newpage
    \,

    \needspace{5\baselineskip}
    \section{Masses at various radii.}\label{appendix:masses}

    Table of masses at $10~{\rm kpc}$ increments out to $250~{\rm kpc}$. Included are the median, the 68th percentile Bayesian credible interval, and the 95th percentile interval. 

    See \url{https://github.com/al-jshen/gmestan-interactive} for an interactive figure showing the full mass distribution and to obtain the mass at arbitrary radii and percentiles. 

    \begin{deluxetable}{CCCCCC}[!ht]
      \tablecaption{Estimated masses at various radii. \label{table:masses}}
      \tablehead{
        \colhead{Radius} & \colhead{2.5th Percentile} & \colhead{16th Percentile} & \colhead{50th Percentile} & \colhead{84th Percentile} & \colhead{97.5th Percentile} \\
        \colhead{kpc} & \multicolumn{5}{c}{$\rm \times 10^{12} \, M_\odot$}
      }
      \startdata
      10  & 0.143 & 0.162 & 0.188 & 0.213 & 0.237 \\
      20  & 0.226 & 0.249 & 0.28  & 0.307 & 0.334 \\
      30  & 0.294 & 0.319 & 0.352 & 0.38  & 0.409 \\
      40  & 0.352 & 0.38  & 0.412 & 0.443 & 0.475 \\
      50  & 0.405 & 0.434 & 0.468 & 0.5   & 0.534 \\
      60  & 0.451 & 0.483 & 0.518 & 0.554 & 0.59 \\
      70  & 0.495 & 0.529 & 0.565 & 0.604 & 0.644 \\
      80  & 0.537 & 0.571 & 0.609 & 0.651 & 0.693 \\
      90  & 0.576 & 0.61  & 0.651 & 0.696 & 0.741 \\
      100 & 0.611 & 0.648 & 0.692 & 0.738 & 0.79 \\
      110 & 0.644 & 0.683 & 0.731 & 0.781 & 0.837 \\
      120 & 0.675 & 0.717 & 0.768 & 0.823 & 0.881 \\
      130 & 0.703 & 0.749 & 0.803 & 0.862 & 0.926 \\
      140 & 0.733 & 0.78  & 0.838 & 0.901 & 0.969 \\
      150 & 0.758 & 0.81  & 0.872 & 0.939 & 1.01 \\
      160 & 0.784 & 0.84  & 0.906 & 0.975 & 1.048 \\
      170 & 0.808 & 0.867 & 0.937 & 1.01  & 1.087 \\
      180 & 0.831 & 0.894 & 0.968 & 1.045 & 1.128 \\
      190 & 0.854 & 0.92  & 0.999 & 1.08  & 1.166 \\
      200 & 0.877 & 0.946 & 1.028 & 1.114 & 1.203 \\
      210 & 0.898 & 0.97  & 1.058 & 1.148 & 1.24 \\
      220 & 0.92  & 0.995 & 1.087 & 1.181 & 1.279 \\
      230 & 0.942 & 1.019 & 1.116 & 1.212 & 1.314 \\
      240 & 0.962 & 1.042 & 1.143 & 1.244 & 1.35 \\
      250 & 0.981 & 1.065 & 1.17  & 1.275 & 1.388 \\
      \enddata
      % \tablecomments{asdf}
    \end{deluxetable}

    \newpage
    \,

    \needspace{5\baselineskip}
    \section{Results from analysis of mock catalogs.}\label{appendix:mocks}

    The table below shows the estimated distribution function parameters, virial radius, and mass enclosed within both $100~{\rm kpc}$ and the virial radius for all runs on simulated galaxies. For each of the four simulated galaxies, there are five runs, one with the full sample (based on our selection criteria; see Section \ref{sec:data}) and four split-sky tests, where for each split-sky test all stars in a quarter of the sky are removed based on right ascension. 

    \begin{deluxetable}{cccccccccc}[!ht]
      \tablecaption{Results from split-sky tests of mock catalogs. \label{table:mocks}}
      \tablehead{
        \colhead{Galaxy} & \colhead{Quadrant removed} & \colhead{Stars} & \colhead{$\alpha$} & \colhead{$\beta$} & \colhead{$\gamma$} & \colhead{$\po$} & \colhead{$r_{200}$} & \colhead{$\rm M(<100 kpc)$} & \colhead{${\rm M_{200}}$} \\
        \colhead{} & \colhead{RA (J2000)} & \colhead{Number} & \colhead{} & \colhead{} & \colhead{} & \colhead{$10^{4}~{\rm km^{2}\,s^{-2}}$} & \colhead{$\rm kpc$} & \colhead{$\times 10^{12}~{\rm M_\odot}$} & \colhead{$\times 10^{12}~{\rm M_\odot}$}
      }
      \startdata
      6 & -180$^\circ$ to -90$^\circ$ removed & 82   & $3.0030_{-0.0022}^{+0.0046}$ & $0.33_{-0.05}^{+0.05}$ & $0.47_{-0.05}^{+0.05}$ & $58.85_{-7.54}^{+7.44}$ & $219.12_{-10.18}^{+10.35}$ & $0.74_{-0.07}^{+0.07}$ & $1.12_{-0.15}^{+0.17}$ \\
      6 & -90$^\circ$ to 0$^\circ$ removed    & 141  & $3.0016_{-0.0012}^{+0.0025}$ & $0.33_{-0.05}^{+0.05}$ & $0.46_{-0.05}^{+0.05}$ & $49.20_{-7.33}^{+7.80}$ & $204.79_{-9.16}^{+9.63}$   & $0.62_{-0.06}^{+0.06}$ & $0.91_{-0.12}^{+0.14}$ \\
      6 & 0$^\circ$ to 90$^\circ$ removed     & 125  & $3.0019_{-0.0014}^{+0.0029}$ & $0.34_{-0.05}^{+0.05}$ & $0.46_{-0.06}^{+0.06}$ & $48.85_{-8.67}^{+8.45}$ & $203.31_{-8.58}^{+9.43}$   & $0.61_{-0.06}^{+0.06}$ & $0.89_{-0.11}^{+0.13}$ \\
      6 & 90$^\circ$ to 180$^\circ$ removed   & 111  & $3.0019_{-0.0015}^{+0.0034}$ & $0.33_{-0.05}^{+0.05}$ & $0.47_{-0.05}^{+0.05}$ & $52.25_{-7.16}^{+7.72}$ & $208.33_{-9.08}^{+10.16}$  & $0.65_{-0.06}^{+0.07}$ & $0.96_{-0.12}^{+0.15}$ \\
      6 & None removed                        & 153  & $3.0013_{-0.0010}^{+0.0022}$ & $0.34_{-0.05}^{+0.05}$ & $0.47_{-0.06}^{+0.06}$ & $48.62_{-7.74}^{+8.39}$ & $203.02_{-9.31}^{+9.88}$   & $0.61_{-0.06}^{+0.07}$ & $0.89_{-0.12}^{+0.14}$ \\
      \hline
      23 & -180$^\circ$ to -90$^\circ$ removed & 70  & $3.0037_{-0.0029}^{+0.0060}$ & $0.29_{-0.05}^{+0.04}$ & $0.46_{-0.05}^{+0.04}$ & $61.03_{-6.71}^{+6.79}$ & $224.98_{-10.14}^{+11.95}$ & $0.78_{-0.07}^{+0.08}$ & $1.21_{-0.16}^{+0.20}$ \\
      23 & -90$^\circ$ to 0$^\circ$ removed    & 88  & $3.0027_{-0.0021}^{+0.0044}$ & $0.29_{-0.04}^{+0.05}$ & $0.46_{-0.05}^{+0.05}$ & $60.26_{-6.98}^{+7.11}$ & $223.96_{-9.69}^{+10.88}$  & $0.77_{-0.07}^{+0.08}$ & $1.20_{-0.15}^{+0.18}$ \\
      23 & 0$^\circ$ to 90$^\circ$ removed     & 63  & $3.0043_{-0.0031}^{+0.0067}$ & $0.30_{-0.05}^{+0.05}$ & $0.45_{-0.05}^{+0.05}$ & $62.30_{-6.86}^{+6.93}$ & $228.79_{-10.65}^{+11.23}$ & $0.81_{-0.08}^{+0.08}$ & $1.27_{-0.17}^{+0.20}$ \\
      23 & 90$^\circ$ to 180$^\circ$ removed   & 70  & $3.0036_{-0.0027}^{+0.0060}$ & $0.29_{-0.05}^{+0.04}$ & $0.46_{-0.05}^{+0.05}$ & $61.71_{-7.08}^{+6.68}$ & $225.08_{-9.58}^{+10.29}$  & $0.78_{-0.07}^{+0.07}$ & $1.21_{-0.15}^{+0.17}$ \\
      23 & None removed                        & 97  & $3.0025_{-0.0018}^{+0.0037}$ & $0.29_{-0.05}^{+0.05}$ & $0.46_{-0.05}^{+0.04}$ & $60.29_{-7.26}^{+6.91}$ & $223.76_{-9.71}^{+10.84}$  & $0.77_{-0.07}^{+0.07}$ & $1.19_{-0.15}^{+0.18}$ \\
      \hline
      24 & -180$^\circ$ to -90$^\circ$ removed & 52  & $3.0057_{-0.0042}^{+0.0112}$ & $0.33_{-0.05}^{+0.05}$ & $0.46_{-0.05}^{+0.05}$ & $60.58_{-7.29}^{+7.07}$ & $222.82_{-9.77}^{+11.70}$  & $0.77_{-0.07}^{+0.08}$ & $1.18_{-0.15}^{+0.20}$ \\
      24 & -90$^\circ$ to 0$^\circ$ removed    & 85  & $3.0028_{-0.0021}^{+0.0042}$ & $0.35_{-0.05}^{+0.05}$ & $0.47_{-0.05}^{+0.05}$ & $59.10_{-7.20}^{+7.43}$ & $219.72_{-10.49}^{+10.53}$ & $0.74_{-0.07}^{+0.07}$ & $1.13_{-0.15}^{+0.17}$ \\
      24 & 0$^\circ$ to 90$^\circ$ removed     & 72  & $3.0038_{-0.0028}^{+0.0053}$ & $0.34_{-0.05}^{+0.05}$ & $0.47_{-0.05}^{+0.05}$ & $59.08_{-7.42}^{+7.19}$ & $219.67_{-10.76}^{+11.16}$ & $0.74_{-0.08}^{+0.08}$ & $1.13_{-0.16}^{+0.18}$ \\
      24 & 90$^\circ$ to 180$^\circ$ removed   & 73  & $3.0033_{-0.0025}^{+0.0056}$ & $0.34_{-0.05}^{+0.05}$ & $0.47_{-0.05}^{+0.05}$ & $59.97_{-6.87}^{+7.35}$ & $221.09_{-10.39}^{+11.25}$ & $0.75_{-0.07}^{+0.08}$ & $1.15_{-0.15}^{+0.18}$ \\
      24 & None removed                        & 94  & $3.0023_{-0.0017}^{+0.0042}$ & $0.35_{-0.05}^{+0.05}$ & $0.47_{-0.05}^{+0.05}$ & $58.96_{-6.92}^{+7.67}$ & $217.71_{-10.11}^{+10.47}$ & $0.73_{-0.07}^{+0.07}$ & $1.10_{-0.15}^{+0.17}$ \\
      \hline
      27 & -180$^\circ$ to -90$^\circ$ removed & 55  & $3.0057_{-0.0043}^{+0.0088}$ & $0.33_{-0.05}^{+0.05}$ & $0.44_{-0.04}^{+0.04}$ & $62.74_{-6.69}^{+6.59}$ & $232.21_{-10.96}^{+10.63}$ & $0.83_{-0.08}^{+0.08}$ & $1.33_{-0.18}^{+0.19}$ \\
      27 & -90$^\circ$ to 0$^\circ$ removed    & 140 & $3.0015_{-0.0012}^{+0.0024}$ & $0.35_{-0.05}^{+0.05}$ & $0.45_{-0.05}^{+0.04}$ & $61.51_{-6.51}^{+6.90}$ & $228.73_{-8.03}^{+10.21}$  & $0.81_{-0.06}^{+0.07}$ & $1.27_{-0.13}^{+0.18}$ \\
      27 & 0$^\circ$ to 90$^\circ$ removed     & 116 & $3.0018_{-0.0013}^{+0.0029}$ & $0.33_{-0.05}^{+0.05}$ & $0.45_{-0.05}^{+0.04}$ & $62.85_{-6.98}^{+6.30}$ & $230.21_{-9.50}^{+9.82}$   & $0.82_{-0.07}^{+0.07}$ & $1.30_{-0.15}^{+0.17}$ \\
      27 & 90$^\circ$ to 180$^\circ$ removed   & 127 & $3.0016_{-0.0012}^{+0.0027}$ & $0.33_{-0.04}^{+0.04}$ & $0.44_{-0.05}^{+0.04}$ & $61.34_{-5.59}^{+6.17}$ & $230.95_{-9.37}^{+10.00}$  & $0.82_{-0.06}^{+0.07}$ & $1.31_{-0.15}^{+0.18}$ \\
      27 & None removed                        & 146 & $3.0013_{-0.0010}^{+0.0023}$ & $0.35_{-0.05}^{+0.05}$ & $0.45_{-0.05}^{+0.04}$ & $61.84_{-6.07}^{+6.58}$ & $229.00_{-8.90}^{+10.28}$  & $0.81_{-0.06}^{+0.07}$ & $1.28_{-0.14}^{+0.18}$ \\
      \enddata
      % \tablecomments{asdf}
    \end{deluxetable}

    \newpage
    \,

    \section{Equations for mass}\label{appendix:mass}

    There is some ambiguity in some of the equations for mass given in the literature with this model. Here we explicitly write out how, given the distribution function used in this paper, one can calculate masses.

    With this model, the mass enclosed within any radius is given by
\begin{align}\label{eq:mass}
  {\rm M}(<r) = \frac{\gamma\,\po\,r_0}{G}\left(\frac{r}{r_0}\right)^{1 - \gamma},
\end{align}
where $\gamma$ is unitless, $\po$ is in units of $10^{4}~{\rm km^{2}\,s^{-2}}$, $r$ is the Galactocentric radius in kpc, $r_0$ is a length scale which we set to be $r_0 = 1~{\rm kpc}$, and $G = 6.67\times 10^{-11}~{\rm m^3\,kg^{-1}\,s^{-2}}$ is the universal gravitational constant.

With $G \equiv 1$ units, this is written as
\begin{align}\label{eq:mass_at_radius_alt}
  \frac{\rm M(<r)}{10^{12}~{\rm M_\odot}} = 2.325 \times 10^{-3}\,\gamma\,\left(\frac{\po}{10^{4}~{\rm km^{2}\,s^{-2}}}\right)\,\left(\frac{r}{1~{\rm kpc}}\right)^{1 - \gamma},
\end{align}
where the resulting $\rm M(<r)$ is the mass of the galaxy within radius $r$ in units of $10^{12}~{\rm M_\odot}$ and $\gamma$, $\po$, and $r$, are the same as in Eq. \ref{eq:mass}.

Alternatively, the circular velocity in $\rm km\,s^{-1}$ is given by
\begin{align}\label{eq:velocity_at_radius}
  V_{circ} = \sqrt{\gamma\,\po\,\left(\frac{r}{r_0}\right)^{-\gamma}},
\end{align}
again with the same definitions of $\gamma$, $\po$, $r$, and $r_0$.

    \newpage
    \,

    \section{The correlation between $\po$ and $\gamma$}\label{appendix:correlation}

    The mass in the model is given by
    \begin{equation}\label{eqn:mass}
      M(<r) = \gamma\frac{\Phi_0}{v_0^2}\left(\frac{r}{r_0}\right)^{1-\gamma}\frac{r_0v_0^2}{G}.
    \end{equation}

    Note that we are being careful with the scale factors $v_0$ and $r_0$; in particular, we are tracking $r_0$, the length that we are using to scale the Galactocentric radii of the stars. We define
    \begin{equation}
      M_0\equiv\frac{r_0v_0^2}{G}.
    \end{equation}

    Taking the natural logarithm of equation \ref{eqn:mass},
    \begin{equation}
      \ln \frac{M(<r)}{M_0} = \ln\gamma+\ln\frac{\Phi_0}{v_0^2}+(1-\gamma)\ln\left(\frac{r}{r_0}\right).
    \end{equation}
    Treating $\Phi_0=\Phi_0(\gamma)$, $M=M(\Phi_0(\gamma),\gamma)$, and taking the derivative of $\ln M/M_0$ with respect to $\gamma$,
    \begin{equation}
      \frac{1}{M}\frac{dM}{d\gamma} = \frac{1}{\gamma}+\frac{1}{\Phi_0}\frac{d\Phi_0}{d\gamma}-\ln\frac{r}{r_0}.
    \end{equation}
    Assuming that the mass is essentially fixed during this variation, i.e., setting the derivative to zero, we find an expression for $d\Phi_0/d\gamma$:
    \begin{equation}\label{eqn:derivative}
      \frac{d\Phi_0}{d\gamma} = \Phi_0\left[\ln\frac{r}{r_0}-\frac{1}{\gamma}\right].
    \end{equation}

    \cite{Deason2021} use length scale $r_0=50$ kpc and have stars between $50$ kpc and $100$ kpc, so $\ln(r/r_0)$ is between 0 and $0.69$. Thus
    \begin{equation}
      \frac{d\Phi_0}{d\gamma}\approx -1.5.
    \end{equation}

    In our analysis, we use $r_0=1$ kpc, with stars around $100$ kpc, so $\ln(r/r_0)\approx4.6$, which is greater than $1/\gamma$, so
    \begin{equation}
      \frac{d\Phi_0}{d\gamma}\approx 3.
    \end{equation}

    Either expression shows that there is likely to be a correlation between $\gamma$ and $\Phi_0$; as long as there is some scatter in one, there will be a correlated scatter in the other, unless one chooses $r_0$ so as to set the right hand side of equation \ref{eqn:derivative} to zero.

    The sign of the correlation depends on the magnitude of the scale factor $r_0$, demonstrating the importance of keeping track of the scale factor when using power law expressions.

\end{CJK*}
\end{document}